\pdfoutput=1

\documentclass[11pt]{article}
\usepackage[utf8]{inputenc}
\usepackage[final]{acl}
\usepackage{times}
\usepackage{latexsym}
\usepackage{amssymb}
\usepackage{graphicx}
\usepackage{newunicodechar}
\newunicodechar{−}{-}
\usepackage{lmodern}
\usepackage{comment}
\usepackage{url}
\usepackage{lettrine}
\usepackage{hyperref}
\usepackage{footnote}
\usepackage{manyfoot}
\usepackage{textgreek}
\usepackage{fancyvrb}
\usepackage{fvextra}
\usepackage{fancyvrb}
\usepackage{enumitem}
\usepackage{hyperref}
\usepackage{multirow}
\usepackage{rotating}
\usepackage{booktabs}

\usepackage{times}
\usepackage{algorithm}
\usepackage{colortbl} 
\usepackage{amsmath}
\usepackage{algorithmic}
\usepackage{subcaption}
\usepackage{todonotes}
\usepackage{amssymb}
\usepackage{textcomp}

\DefineVerbatimEnvironment{CustomVerbatim}{Verbatim}{fontsize=\small, frame=single, breaklines=true, breaksymbolleft={}}

\usepackage{microtype}

\usepackage{inconsolata}
\usepackage{booktabs}
\usepackage{multirow}
\usepackage{graphicx}

\usepackage{colortbl}

\title{CRScore++: Reinforcement Learning with Verifiable Tool and AI Feedback for Code Review}

\author{Manav Nitin Kapadnis\footnotemark[1] ~~ Atharva Naik\footnotemark[1] ~~ Carolyn Ros\'e \\
  Language Technologies Institute \\
  Carnegie Mellon University \\
  \texttt{\{mkapadni, arnaik, cprose\}@cs.cmu.edu}}

\begin{document}
\maketitle
\begin{abstract}

Reinforcement learning (RL) to improve code review comment generation requires handling unstructured outputs, making reinforcement learning (RL) feedback challenging. The two main RL approaches, namely RL with Verifiable Feedback (RLVR) and RL with AI Feedback (RLAIF), offer trade-offs: RLVR provides reliable feedback for structured tasks like code generation, while RLAIF works for unstructured outputs but is subjective. We bridge this gap with CRScore++, an RL framework that leverages both LLM-based subjective feedback and verifiable signals for training. Extending CRScore, a code review evaluation metric integrating LLMs with verifiers like linters and code smell detectors, CRScore++ transforms these signals into training rewards. We show that CRScore++ improves a weaker student model through a combination of supervised fine-tuning and RL critique from a stronger teacher model, thus enabling generalization to novel programming languages.
\end{abstract}

\renewcommand{\thefootnote}{\fnsymbol{footnote}}
\footnotetext[1]{Equal contribution.}









\section{Introduction}
\label{Chapter 1}

In recent years, large language models (LLMs) have been increasingly adopted for software engineering tasks \cite{zhang2023unifying}. This trend has been driven in part by advancements in reinforcement learning (RL)-based training methods, particularly those incorporating verifiable rewards (RLVR), such as execution feedback \cite{codeRL, dou2024stepcoderimprovecodegeneration, shojaee2023executionbasedcodegenerationusing}. While these learning signals have proven effective for tasks with structured outputs like code generation, applying verifiable rewards to natural language generation (NLG) tasks, such as code review, remains challenging.

For NLG tasks, the research community has primarily relied on supervised fine-tuning (SFT) with high-quality examples, RL from human feedback (RLHF) \cite{rlhf}, or RL from AI feedback (RLAIF) \cite{rlaif} for greater scalability. While RLVR is both scalable and verifiable, RLHF and RLAIF depend on subjective preference signals \cite{openai2024gpt4, bai2022constitutionalaiharmlessnessai}, which can introduce unintended biases into models. On the other hand, some of these biases may be desirable, as evidenced by studies on using LLMs as evaluators (a form of RLAIF) for software engineering tasks, which demonstrate high agreement with human evaluators and similar score distributions \cite{he2025codecourtroomllmsnew, wang2025llmsreplacehumanevaluators}.

Applying RLVR to code review is particularly difficult due to the open-ended nature of textual outputs. As a result, recent work has explored integrating LLMs with verifiable static analysis tools, such as linters and code smell detectors, either during inference or for data generation. These approaches have been shown to improve both the evaluation \cite{crscore_paper} and generation \cite{jaoua2025combininglargelanguagemodels} of review comments. Building on these ideas, we take steps toward leveraging verifiers in training by incorporating verifiable ``facts'' from static analysis tools alongside more subjective LLM feedback (RLAIF). This results in a new training framework, which we call CRScore++.

Our framework consists of two stages: (1) learning from demonstration (SFT) and (2) learning from critique (Preference Optimization or PO), both of which incorporate verifier-generated ``facts'' into step-by-step reasoning. In the SFT stage, a stronger teacher model generates code review demonstrations for a student model using a chain-of-thought (CoT) approach that integrates relevant linter messages and code smells into the review demonstration. This builds on recent findings that show how incorporating external tools enhances the reasoning abilities of long CoT models \cite{li2025startselftaughtreasonertools}. In the PO stage, the teacher model critiques the student model’s review comments by first generating a list of expected review points (pseudo-references generated by CRScore \cite{crscore_paper}), incorporating tool outputs, and then applying CoT reasoning to evaluate whether these points are addressed. 

Following the findings of \cite{sharma2024criticalevaluationaialign}, we use the same teacher model for both the SFT and PO stages\footnote{To reduce OpenAI API usage costs, we distill the target model into an open-source reward model for the PO stage}. Beyond grounding RLAIF with verifiable facts from tools, our framework also indirectly trains the model to detect code smells and linter-related issues through CoT reasoning. 
A key advantage of our approach is its remarkable cross-language generalization capability. Models trained exclusively on Python code achieve nearly equivalent performance when generating reviews for Java and JavaScript, despite having no exposure to these languages during training. This result challenges the prevailing assumption that language-specific training data is essential for high-quality code review systems.

Our work makes three primary contributions:
\begin{itemize}[noitemsep,topsep=0pt,leftmargin=*]
    \item Introducing a novel RL method for training code models that combines the benefits of RLVR with the flexibility of RLAIF.
    \item Applying this framework to integrate linters, code smell detectors, and security analyzers to measure review quality for code model training, which to the best of our knowledge hasn't been explored in prior work.
    \item Enabling generalizable learning of skills such as code smell detection, code efficiency evaluation, and assessments of security, readability, and maintainability through step-by-step reasoning.
    \item Demonstrating robust cross-language generalization of code quality analysis abilities. Models trained exclusively on Python data maintain most of their performance when reviewing Java and JavaScript code without any language-specific fine-tuning.
\end{itemize}
\section{Related Work}
\label{Chapter 2}

\subsection{Code Review Automation}

Recent work on automated code review leverages large language models (LLMs) to generate reviews that resemble human-written reviews \cite{codereviewerpaper}. While systematic assessment can be done using benchmarking datasets \cite{tufano}, conventional metrics like BLEU fail to measure semantic quality of generated reviews. Prior work demonstrated the potential of LLMs as automatic code reviewers \cite{he2025code} and identified key reasoning abilities needed for code review comprehension \cite{lin2025codereviewqa}, but less attention has been paid to evaluating review comment quality.
Hybrid approaches for review comment generation combining static analyzers and LLMs \cite{jaoua2025combining} are promising but concentrate largely on training data or inference time augmentation and, to a lesser extent, on the utilization of verification signals as rewards during training. 
\citet{wang2025llmjudge, jaoua2025combining} and \citet{crscore_paper} all show that LLM-as-a-judge methods achieve a significant correlation with human scores for software engineering and code review evaluation. 
Moreover, CRScore \cite{crscore_paper} shows that integrating static analysis tools with LLMs and using pseudo-references can lead to a more accurate signal of review quality.
Based on past evidence of how LLMs combined with tools can provide strong signals of review quality, our work extends this idea by using them as a reward signal for training LLMs to generate better review comments.
Additionally, we also show that supervision and knowledge obtained from a tool for one language (e.g. Python) can lead to better reviews for unseen languages at test time (e.g. Java and Javascript).

\subsection{Static Analysis Integration}

Static analysis tools provide structured signals regarding code quality by detecting code smells and enforcing style and maintainability standards \cite{Pecorelli2022-bp}. Early work limited integration with LLMs to post-generation validation, but recent research has explored deeper synergy. \citet{combining_LLM_with_static_analyzer} systematically combined static analyzers and LLMs at multiple stages, including data augmentation, retrieval-augmented generation, and naive output concatenation, thus demonstrating that hybrid strategies can improve review relevance and coverage. However, their evaluation primarily relies on LLM as a judge relative comparison, whereas our approach adapts CRScore \cite{crscore_paper} with pseudo-references, which was shown to be more scalable and aligns better with human preferences. Additionally, we also incorporate code smell detectors in addition to the linters and integrate them into the reasoning process of the model, thus showing that tool-based knowledge learned in one language can generalize to others.

ReTool \cite{feng2025retoolreinforcementlearningstrategic} further motivates our methodology by showing that reinforcement learning over tool feedback can enhance strategic reasoning and tool utilization. We extend these ideas by integrating tool signals directly into both the demonstration and preference optimization stages of training, enabling our models to leverage static analysis feedback for both review generation and cross-language generalization.

\subsection{Reinforcement Learning from Subjective Feedback}

Preference-based alignment methods like RLHF \cite{rlhf} and RLAIF \cite{rlaif} face intrinsic limitations when applied to specialized domains that require objective assessments of quality \cite{RLVR}. Although such methods improve results in subjective settings \cite{sharma2022surveymachinelearningtechniques}, their reliance on the preference of human or artificial intelligence results in biases that conflict with best practices in software engineering \cite{mcaleese2024llmcritics}. The challenge is compounded in code review settings, where the produced feedback must balance technical correctness with readability for human readers. Our suggested framework resolves this tension through the combination of verifiable signals from static analysis with preference-based learning, thus creating a hybrid reward system grounded in both objective metrics and linguistic quality.

\subsection{Reinforcement Learning from Verifiable Rewards}

The RLVR framework has proven to be effective in code generation \cite{chen2022codetcodegenerationgenerated} and mathematical reasoning domains \cite{shao2024deepseekmathpushinglimitsmathematical}, where its execution stage provides clear verification signals. Recent advances in multi-step reinforcement learning \cite{goldie2025synthetic} and process supervision \cite{dai2025processsupervisionguidedpolicyoptimization} describe how intermediate-level verification can promote sophisticated reasoning abilities. The curriculum learning approach utilized by StepCoder \cite{dou2024stepcoderimprovecodegeneration} incorporates segmentation of code review into verifiable subtasks. CRScore++ applies these concepts to natural language generation by leveraging static analyzer outputs as partial verifiers, thus providing objective reward signals and at the same time solving for contextual feedback generation requirements. The process described here solves verification issues common for natural language generation settings while at the same time benefiting from the learned linguistic structures. 
\section{Methodology} 
\label{Chapter 4}

\begin{figure*}[!tbh]
\centering
\includegraphics[width=1.6\columnwidth]{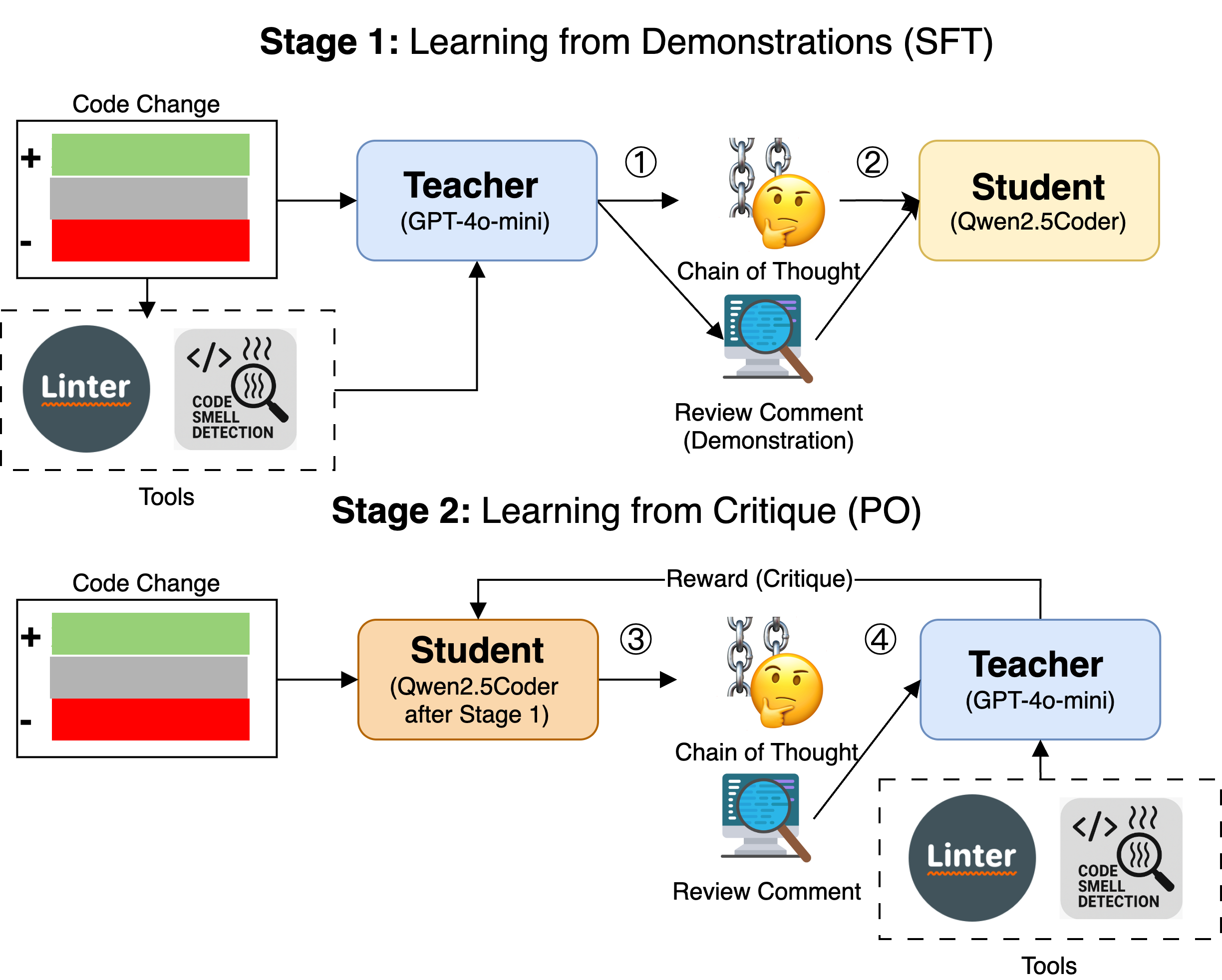}
\caption{Three-stage training pipeline: (A) Tool-augmented distillation, (B) Preference ranking, (C) Cross-language evaluation}
\label{fig:training_stages}
\end{figure*}

In this section, we elaborate on our three-stage methodology for aligning code review generation with verifiable static analysis signals: (1) tool-augmented knowledge distillation, (2) direct preference optimization, and (3) cross-language evaluation on in-domain and out-of-domain coding languages' samples. Figures~\ref{fig:training_stages} which depicts our training workflow and~\ref{fig:evaluation_llm_as_a_judge} shows our evaluation framework, illustrate the complete pipeline.

\subsection{Knowledge Distillation via Demonstration}
We distill GPT-4o Mini's \cite{gpt4o-mini} code review capabilities into Qwen2.5-Coder-Instruct \cite{hui2024qwen2_5} models (3B and 7B models) using structured prompts (Appendix \ref{app:sft_dataset_creation_prompt}) that integrate static analysis feedback. For each Python code change in the CodeReviewer dataset \cite{codereviewerpaper}, we generate linter feedback using Ruff\footnote{\url{https://beta.ruff.rs/}} for identifying code efficiency issues, complexity measurements, and security vulnerabilities
and code smells using Pyscent\footnote{\url{https://github.com/whyjay17/Pyscent}} detecting patterns like Long Method and Large Class amongst many other smells.

Specifically, the teacher model processes these inputs through a structured prompt (Appendix \ref{app:training_prompt}) requiring generation of a step-by-step analysis followed by concise review generation comprising of these code smells and linter feedback signals. As shown in Figure~\ref{fig:training_stages} (Stage 1), this stage combines code changes with tool outputs to facilitate the teacher model's generation of high-quality demonstrations. We train student models using standard cross-entropy loss over 20,888 Python samples for 2 epochs on 2 NVIDIA A100 80GB GPUs with a learning rate of 2e-05 using the Adam \cite{adam_paper} optimizer.

\subsection{Direct Preference Optimization}
In the second stage of our training pipeline as shown in Figure~\ref{fig:training_stages} (Stage 2), we enhance alignment through preference learning using 5,000 Python code changes by sampling 20 candidate reviews per sample using SFT models. These are then scored by GPT4o-mini, which generates a step-by-step analysis and final review scores over a scale of 5 points on comprehensiveness, conciseness, and relevance metric. We then create preference pairs \((y_w,y_l)\) with \(\Delta_{\text{score}} \geq 2\) for alignment training of DPO model.

The DPO objective \cite{dpo} optimizes:

\begin{equation} 
\resizebox{ \columnwidth}{!}{$
\mathcal{L}_{\text{DPO}} = -\mathbb{E}\left[\log\sigma\left(\beta\log\frac{\pi_\theta(y_w|x)}{\pi_{\text{ref}}(y_w|x)} - \beta\log\frac{\pi_\theta(y_l|x)}{\pi_{\text{ref}}(y_l|x)}\right)\right]$
}\end{equation}
\\
with \(\beta=0.1\), which balances preference adherence against regularization. Training runs for 2 epochs using the Adam optimizer \cite{adamw} with 2e-5 learning rate.

\subsection{Cross-language Evaluation}
\begin{figure*}[!tbh]
\centering
\includegraphics[width=1.75\columnwidth]{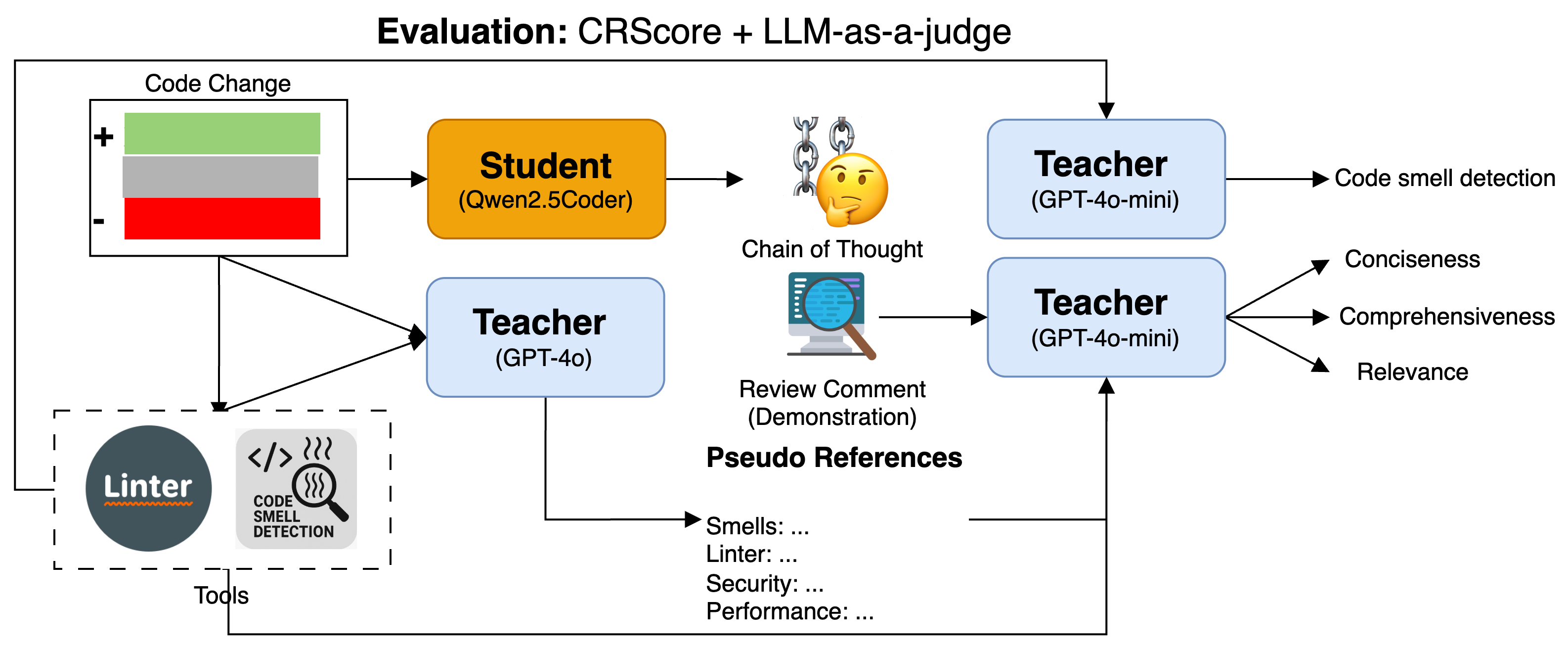}
\caption{Evaluation framework combining static verification with LLM-as-a-judge}
\label{fig:evaluation_llm_as_a_judge}
\end{figure*}

We test generalization by using Python alongside Java and JavaScript code changes in the test sets from the CodeReviewer dataset as shown in Figure~\ref{fig:evaluation_llm_as_a_judge}, using the respective programming language-specific tools for evaluation. For Java code patch analysis, we use PMD\footnote{\url{https://pmd.github.io/}} as a static code analysis tool and DesigniteJava \cite{designite} for detecting code smells. We used PMD together with JSNOSE \cite{jsnose} to develop an adapted analyzer for detecting code smells for JavaScript. It is capable of analyzing files in JavaScript for both linter violations and code smells.

The framework used for the evaluation is a two-phase process. First, the GPT-4o Mini creates a comprehensive list of topics considered crucial to evaluate a model with regard to changes made to the code. The model is prompted to identify key topics like code smells, linter warnings, security vulnerabilities, compatibility issues, readability issues, performance impact, design patterns, and memory-related issues. The process creates pseudo ground-truths for evaluating reviews generated by the models.

In the second step, our trained models (SFT and SFT+DPO) generate step-by-step analysis and final reviews for all test samples. The final reviews are then compared with the ground truths of the pseudo-topics, code changes made, and feedback provided by the tool, and are evaluated by GPT-4o-Mini. The rating prompt (Appendix \ref{app:llm_as_a_judge_prompt}) asks for an overall review to identify whether all required topics have been included and to provide three different scores:

\begin{itemize}

\item \textbf{Comprehensiveness} (1-5): How well does this review cover all relevant subjects?
\item \textbf{Conciseness} (1-5): Whether the review is appropriately focused without redundancy
\item \textbf{Relevance} (1-5): The degree to which issues specified are related to the code modification.
\end{itemize}

The LLM-as-a-judge framework allows for consistent review across different languages, despite differences in their static tool chains.

In cross-language generalization analysis, we compare how models trained solely with Python instances perform when evaluating Java and JavaScript code changes. This evaluation assesses the models' ability to transfer knowledge about code quality principles across language boundaries. Unlike previous approaches, such as that of \citet{combining_LLM_with_static_analyzer}, which primarily trains and evaluates within the same language using data augmentation techniques (DAT), our methodology specifically tests whether the implicit knowledge about good coding practices, security vulnerabilities, and code smells learned from Python static analyzers during training can be effectively applied to entirely different programming languages at inference time without additional language-specific fine-tuning.

\section{Results} 
\label{Chapter: 6_experimental} 

\subsection{In-Domain Performance}
Table~\ref{tab:python_results} presents evaluation results for code review generation on Python code changes. We observe that models trained with our CRScore++ approach significantly outperform all baseline configurations across comprehensiveness and relevance metrics. The progression from Zero Shot to Stage 1 to Stage 2 shows consistent improvements in these dimensions for both model sizes. Notably, models trained directly on CodeReviewer data without our approach perform worse than even Zero Shot configurations. A clear trade-off emerges between comprehensiveness/relevance and conciseness, as DPO-trained models generate more thorough reviews at the expense of brevity. The 7B variants generally outperform their 3B counterparts, with the Stage 2 7B model achieving the highest comprehensiveness and relevance scores among all configurations.

\begin{table}[htbp]
\centering
\resizebox{\columnwidth}{!}{%
\begin{tabular}{llll}
\toprule
\textbf{Model} & \textbf{Compre.} & \textbf{Conci.} & \textbf{Relev.} \\
\midrule
Qwen 3B (Zero Shot)                 & 0.43 & 0.67 & 0.45 \\
Qwen 3B (CR dataset\textsubscript{\texttt{Python only}})   & 0.10 {\color{red}(-0.33)} & 0.58 {\color{red}(-0.09)} & 0.20 {\color{red}(-0.25)} \\
Qwen 3B (Tool Guided)               & 0.47 {\color{green}(+0.04)} & 0.67 (0.00)              & 0.47 {\color{green}(+0.02)} \\
Qwen 3B (Stage 1)                   & 0.59 {\color{green}(+0.16)} & 0.67 (0.00)              & 0.58 {\color{green}(+0.13)} \\
Qwen 3B (Stage 2)                   & 0.67 {\color{green}(+0.24)} & 0.59 {\color{red}(-0.08)} & 0.64 {\color{green}(+0.19)} \\
\midrule
Qwen 7B (Zero Shot)                 & 0.41 & 0.62 & 0.43 \\
Qwen 7B (CR dataset\textsubscript{\texttt{Python only}})   & 0.12 {\color{red}(-0.29)} & 0.56 {\color{red}(-0.06)} & 0.22 {\color{red}(-0.21)} \\
Qwen 7B (Tool Guided)               & 0.45 {\color{green}(+0.04)} & 0.64 {\color{green}(+0.02)} & 0.46 {\color{green}(+0.03)} \\
Qwen 7B (Stage 1)                   & 0.56 {\color{green}(+0.15)} & 0.68 {\color{green}(+0.06)} & 0.55 {\color{green}(+0.12)} \\
Qwen 7B (Stage 2)                   & 0.69 {\color{green}(+0.28)} & 0.57 {\color{red}(-0.05)} & 0.65 {\color{green}(+0.22)} \\
\bottomrule
\end{tabular}%
}
\caption{Results on in-domain Python language.  All values rounded to two decimals.  Change versus the Zero Shot model shown in parentheses. \textbf{Green “($+$x.xx)” indicates an increase}, \textbf{red “($−$x.xx)” indicates a decrease}.}
\label{tab:python_results}
\end{table}

\subsection{Cross-Language Generalization}
The results for Java and JavaScript code review generation, shown in Table~\ref{tab:java_js_results}, demonstrate strong generalization capabilities despite training exclusively on Python data. Both SFT and DPO models maintain similar performance patterns across languages, with Stage 2 models consistently achieving the highest comprehensiveness and relevance scores for all languages. This robust cross-language transfer is particularly noteworthy as it occurs without any language-specific fine-tuning, suggesting our approach captures fundamental code quality principles rather than language-specific patterns. The relative performance ranking of different approaches remains consistent across languages, suggesting that the benefits of our training methodology transfer effectively to unseen programming languages. We observe that the performance gap between in-domain (Python) and out-of-domain (Java/JavaScript) is minimal for our CRScore++ models, while the gap is more pronounced for zero-shot approaches. 

\begin{table*}[!htbp]
\centering
\resizebox{1.9\columnwidth}{!}{%
\begin{tabular}{lllllll}
\toprule
\multirow{2}{*}{\textbf{Model}} 
  & \multicolumn{3}{c}{\textbf{Java}} 
  & \multicolumn{3}{c}{\textbf{JavaScript}} \\
\cmidrule(lr){2-4}\cmidrule(lr){5-7}
  & \textbf{Compre.} & \textbf{Conci.} & \textbf{Relev.}
  & \textbf{Compre.} & \textbf{Conci.} & \textbf{Relev.} \\
\midrule
Qwen 3B (Zero Shot)                  & 0.43 & 0.67 & 0.46 & 0.42 & 0.66 & 0.44 \\
Qwen 3B (Tool Guided)       & 0.39 {\color{red}(-0.04)}& 0.67 (0.00)               & 0.42 {\color{red}(-0.04)} & 0.39 {\color{red}(-0.03)} & 0.67 {\color{green}(+0.01)}& 0.42 {\color{red}(-0.02)} \\
Qwen 3B (CR dataset\textsubscript{\texttt{Python only}})    & 0.12 {\color{red}(-0.31)}& 0.58 {\color{red}(-0.09)} & 0.22 {\color{red}(-0.24)} & 0.12 {\color{red}(-0.30)} & 0.58 {\color{red}(-0.08)} & 0.22 {\color{red}(-0.22)}\\
Qwen 3B (Stage 1)                    & 0.56 {\color{green}(+0.13)} & 0.67 {\color{red}(-0.01)}  & 0.55 {\color{green}(+0.10)} & 0.53 {\color{green}(+0.11)} & 0.68 {\color{green}(+0.01)}  & 0.53 {\color{green}(+0.09)} \\
Qwen 3B (Stage 2)                    & 0.64 {\color{green}(+0.21)} & 0.61 {\color{red}(-0.06)}& 0.61 {\color{green}(+0.15)} & 0.62 {\color{green}(+0.20)} & 0.61 {\color{red}(-0.06)}  & 0.60 {\color{green}(+0.16)}\\
\midrule
Qwen 7B (Zero Shot)                  & 0.40 & 0.60 & 0.41 & 0.39 & 0.60 & 0.41 \\
Qwen 7B (Tool Guided)       & 0.38 {\color{red}(-0.02)} & 0.61 {\color{green}(+0.01)}& 0.40 {\color{red}(-0.01)} & 0.37 {\color{red}(-0.02)} & 0.59 {\color{red}(-0.01)} & 0.38 {\color{red}(-0.03)} \\
Qwen 7B (CR dataset\textsubscript{\texttt{Python only}})    & 0.14 {\color{red}(-0.26)} & 0.59 {\color{red}(-0.01)} & 0.24 {\color{red}(-0.17)} & 0.15 {\color{red}(-0.24)} & 0.58 {\color{red}(-0.02)} & 0.24 {\color{red}(-0.17)}\\
Qwen 7B (Stage 1)                    & 0.54 {\color{green}(+0.14)} & 0.68 {\color{green}(+0.08)} & 0.53 {\color{green}(+0.12)} & 0.51 {\color{green}(+0.12)} & 0.68 {\color{green}(+0.08)} & 0.52 {\color{green}(+0.11)} \\
Qwen 7B (Stage 2)                    & 0.67 {\color{green}(+0.27)} & 0.57 {\color{red}(-0.03)}& 0.63 {\color{green}(+0.22)} & 0.65 {\color{green}(+0.26)} & 0.58 {\color{red}(-0.02)}  & 0.62 {\color{green}(+0.21)} \\
\bottomrule
\end{tabular}%
}
\caption{Results on out-of-domain Java and JavaScript. All values rounded to two decimals. In parentheses: absolute change versus the Zero Shot model. The Zero Shot rows omit change annotations. \textbf{Green “(+x.xx)” indicates an increase, red “(−x.xx)” indicates a decrease versus zero‐shot.}}
\label{tab:java_js_results}
\end{table*}

\subsection{Tool Utilization Analysis}

Table~\ref{tab:tool_usage_accuracy_combined} examines how effectively models use structured prompts (Appendix \ref{app:cot_evaluation_prompt}) to incorporate feedback from static analysis tools. We observe that accuracy scores remain relatively stable across different configurations, suggesting that models maintain factual correctness when referencing tool output. However, coverage scores show a marked improvement from zero shot with tools to the SFT and DPO stages, indicating that our training approach helps models better utilize available tool feedback. The 3B DPO model achieves the highest coverage score, showing that smaller models can effectively learn to incorporate tool signals through our training methodology. Although 7B models show slightly higher accuracy in the SFT stage, the difference in performance between model sizes is less pronounced for tool utilization than for overall review quality metrics.


\begin{table*}[htbp]
\centering
\resizebox{2\columnwidth}{!}{
\begin{tabular}{lcccccc}
\toprule
\textbf{Model Name}  
  & \multicolumn{2}{c}{\textbf{Python}} 
  & \multicolumn{2}{c}{\textbf{Java}} 
  & \multicolumn{2}{c}{\textbf{JavaScript}} \\
\cmidrule(lr){2-3}\cmidrule(lr){4-5}\cmidrule(lr){6-7}
  & \textbf{Avg Accuracy} & \textbf{Avg Coverage} 
  & \textbf{Avg Accuracy} & \textbf{Avg Coverage} 
  & \textbf{Avg Accuracy} & \textbf{Avg Coverage} \\
\midrule
Qwen 3B (Zero Shot)  & 2.26 & 1.55 & 2.63 & 1.47 & 2.68 & 1.11 \\
Qwen 3B (Stage 1)    & 3.38 & 1.90 & 3.29 & 1.88 & 3.72 & 2.35 \\
Qwen 3B (Stage 2)    & 2.97 & 1.62 & 3.07 & 1.72 & 3.02 & 1.42 \\
\midrule
Qwen 7B (Zero Shot)  & 2.61 & 1.61 & 3.04 & 1.60 & 3.03 & 1.14 \\
Qwen 7B (Stage 1)    & 3.59 & 1.93 & 3.50 & 1.84 & 3.71 & 2.03 \\
Qwen 7B (Stage 2)    & 3.29 & 2.01 & 3.12 & 1.44 & 3.20 & 1.80 \\
\bottomrule
\end{tabular}}
\caption{Tool usage accuracy and coverage evaluation on Python, Java, and JavaScript using GPT4o-mini as judge.}
\label{tab:tool_usage_accuracy_combined}
\end{table*}

\section{Discussion} 
\label{Chapter 6_discussion} 

\subsection{In-Domain Performance Analysis}
Our experiments reveal several counterintuitive findings about code review training paradigms. Surprisingly, models trained directly on CodeReviewer's ``ground truth'' (scraped from online code review discussions) perform worse than zero-shot approaches. 
This validates the observation from CRScore \cite{crscore_paper} that review comments scraped from online discussions are often noisy, irrelevant. 
The Tool Guided setting demonstrates improved comprehensiveness compared to zero-shot, highlighting the value of integrating static analysis results. 
Our CRScore++ methodology addresses these limitations through a two-stage approach. The SFT stage (Stage 1) achieves a balance between comprehensiveness, relevance, and conciseness, integrating tool feedback while maintaining review brevity. 
The DPO stage (Stage 2) further enhances comprehensiveness and relevance at the expense of some conciseness, suggesting that preference optimization prioritizes thorough coverage of issues over brevity. 
This comprehensiveness-conciseness tradeoff offers a practical advantage, allowing practitioners to select either SFT models (for balanced reviews) or DPO models (for comprehensive analysis) based on specific requirements.

\subsection{Cross-Language Generalization Patterns}
The strong out-of-domain performance (Table~\ref{tab:java_js_results}) challenges assumptions about language-specific code quality analysis. Models trained with CRScore++ exclusively on Python achieve comparable performance on reviewing Java and JavaScript code changes, indicating they learn general code quality analysis skills about security, maintainability, and efficiency rather than language-specific heuristics. The consistent performance hierarchy across languages, with DPO models maintaining the smallest generalization gap, suggests that preference optimization enhances the transfer of fundamental software engineering knowledge. This finding has significant practical implications, as it enables the deployment of unified code review systems in polyglot codebases without language-specific training.

\subsection{Tool Interaction Dynamics}
The tool usage analysis (Table~\ref{tab:tool_usage_accuracy_combined}) reveals three key insights: (1) Model scale (3B vs 7B) has negligible impact on tool utilization accuracy, (2) DPO provides no accuracy improvements over SFT, indicating tool understanding is primarily established during initial distillation, (3) Coverage improvements from SFT/DPO stages demonstrate models learn to surface more tool-identified issues without hallucination. These patterns imply that while preference optimization refines review quality, the core ability to interpret and apply tool feedback is solidified during supervised fine-tuning. 
The stability of accuracy scores across configurations further validates that tool grounding prevents factual errors in generated reviews.

\subsection{Human Study}
\label{sec:human_study}




\begin{table*}[htbp]
\centering
\resizebox{1.75\columnwidth}{!}{%
\begin{tabular}{lcccccccccc}
\toprule
\multirow{2}{*}{\textbf{Model}} & \multicolumn{3}{c}{\textbf{Python}} & \multicolumn{3}{c}{\textbf{Java}} & \multicolumn{3}{c}{\textbf{JavaScript}} \\
\cmidrule(lr){2-4}\cmidrule(lr){5-7}\cmidrule(lr){8-10}
 & \textbf{Comp.} & \textbf{Conc.} & \textbf{Rel.} & \textbf{Comp.} & \textbf{Conc.} & \textbf{Rel.} & \textbf{Comp.} & \textbf{Conc.} & \textbf{Rel.} \\
\midrule
3B (Zero Shot) & 2.79 & \textbf{3.85} & 2.77 & 2.90 & 3.84 & 2.99 & 3.00 & 3.73 & 3.18 \\
7B (Zero Shot) & - & - & - & 2.97 & 3.59 & 2.81 & 3.21 & \textbf{3.92} & 3.21 \\
3B (Tool Guided) & 3.15 & 3.53 & 3.08 & 2.94 & 3.61 & 2.83 & 2.82 & 3.52 & 2.79 \\
7B (Tool Guided) & - & - & - & 2.93 & \textbf{4.04} & 3.05 & 2.65 & 3.42 & 2.82 \\
3B (Stage 1) & \underline{3.56} & \underline{3.71} & \underline{3.17} & 3.76 & \underline{3.88} & \textbf{3.46} & \underline{3.72} & \underline{3.81} & \underline{3.44} \\
7B (Stage 1) & - & - & - & 3.45 & 3.76 & 3.11 & 3.39 & 3.76 & 3.26 \\
3B (Stage 2) & \textbf{3.91} & 3.26 & \textbf{3.19} & \underline{3.88} & 3.42 & 3.19 & 4.09 & 3.38 & \textbf{3.51} \\
7B (Stage 2) & - & - & - & \textbf{4.39} & 3.18 & \underline{3.41} & \textbf{4.12} & 3.24 & 3.48 \\
\bottomrule
\end{tabular}%
}
\caption{Human evaluation results across Python, Java, and JavaScript (scores range 1-5, higher is better). Best scores are \textbf{bolded} and second-best are \underline{underlined}.}
\label{tab:human_eval}
\end{table*}

To help ensure the validity of our automated evaluation framework, we conducted a human evaluation study on 100 code changes (33 per programming language: Python, Java, and JavaScript). 
Human raters conducted evaluations on code reviews generated from four different model configurations: Zero Shot, Tool Guided, Stage 1 (SFT), and Stage 2 (DPO). For the Python programming language, we limited our evaluation to 3B models, based on the consistent performance trends observed across model sizes in the automated metrics.
The set up of the study was inspired by the review quality dimension validity study from CRScore \cite{crscore_paper}.
Annotators rated each review on a 5-point Likert scale across three dimensions: comprehensiveness (coverage of relevant issues), conciseness (appropriate focus), and relevance (pertinence to code changes). We observed moderate inter-annotator agreement (0.65), indicating reliable consensus among evaluators.

Table~\ref{tab:human_eval} shows the results of human evaluations, which follow the same trends as the automated evaluation. 
Stage 2 (DPO) models achieve the top scores for comprehensiveness on all the metrics, with the 7B version showing particularly high expertise on Java (4.39) and JavaScript (4.12). Zero-shot approaches show strong performance on the conciseness dimension, however, they lack comprehensiveness and relevance. 
In contrast, Stage 1 (SFT) models show a balanced level of performance, often achieving second-best scores across most metrics, particularly in tests with Python.
The human subject empirical study supports the comprehensiveness-conciseness trade-off of automated evaluation metrics. 
In particular, Stage 2 models show reduced conciseness while showing increased comprehensiveness. Most importantly, the results validate the capacity of our method to generalize to other languages. Models trained exclusively with Python data repeatedly show strong performance trends when evaluated on Java and JavaScript code reviews. 
This attests that CRScore++ effectively transfers knowledge of code quality to other programming languages. More elaborate details of the human study guidelines are available in the Appendix (\ref{sec:codebook_for_rating_code_review_quality} and \ref{sec:rating_review_quality_using_topics})
 
\section{Conclusion} 
\label{sec:conclusion}

In this work, we present CRScore++, a novel system that successfully closes the gap between verifiable rewards and artificial intelligence feedback for natural language code review generation. 
Our method shows that combining static analysis tools with LLM judges can produce partial verifiers that help align LLM outputs with quality dimensions in domains lacking perfect verification.
Through careful empirical evaluation, we have shown that CRScore++ achieves significant improvements along three critical axes: (1) improved quality of code reviews, with a 56\% relative improvement in comprehensiveness for 3B models and a 42\% relative improvement in relevance compared to zero-shot baselines, (2) the emergence of intrinsic code smell and code quality analysis capabilities, and (3) generalizability across programming languages, with Python-trained models being able to generate high quality reviews for Java and Javascript and also generalizing abilities like code smell and code quality analysis without explicit training.

The results of our research highlight the importance of a two-stage training approach that first establishes foundational knowledge through supervised fine-tuning on tool-enhanced demonstrations, followed by preference optimization to enhance key quality aspects. The fact that models trained directly on human reviews perform worse than the untrained base LLMs points to the importance of incorporating objective quality metrics into training. 
Furthermore, the small performance gap between in-domain and out-of-domain languages suggests that our approach equips the LLM with fundamental code quality analysis skills, enabling cross-lingual generalization.


\section{Future Work} 
\label{sec:future_work} 


This research proposes several promising avenues for future research efforts. First, examining cross-paradigm transfer in clearly diverse programming paradigms (e.g., functional programming languages represented by Scala and imperative programming languages by Python) would increase understanding of more universal principles of code quality. 
Next, applying Group Relative Policy Optimization (GRPO) could provide more robust generalization of code quality analysis skills across different programming languages. 
Finally, designing reward functions that balance the trade-off between comprehensiveness and conciseness without introducing length bias could further improve our approach across diverse code review scenarios, including repository-level reviews.

Other directions include the integration of additional static analysis techniques, such as runtime performance profilers and dependency analysis tools, to increase the verifiable signal space, investigation of the application of our partial verification technique to other resource-limited technical natural language generation projects, e.g., API documentation or security vulnerability reporting, and investigation of the variation of distillation effectiveness with respect to model size and architecture, which can give useful insights on resource-constrained applications in real-world settings, especially in scenarios where the availability of computational resources can limit the use of large models.

\section{Limitations} 
\label{sec:limitations} 
First, while our human evaluation protocol is methodologically sound, it was conducted with a relatively small number of data points (33 cases per language), which can limit the discovery of edge cases in real-world code review use. Second, the use of GPT-4o-mini as an evaluation metric relies on proprietary infrastructure that can have hidden biases. However, we mitigate this by cross-checking results with human evaluation. 
Third, since the current implementation supports only three programming languages: Python, Java and JavaScript, there are questions about its relevance to resource-limited or domain-specific languages.

The experimental design used here has some limitations for architectural exploration since it focused exclusively on the Qwen model family, and this might limit the insights into the validity of trends across model families. 
We also chose not to use long chain-of-thought (LCoT) models as teacher or reward models in our framework to reduce cost and training time, despite their potential for better tool knowledge utilization.
Finally, our static analysis toolkit, while covering common code smells and linter violations, does not address more advanced aspects of software quality like test coverage, architectural consistency, and comprehensively modeling security threats. 
The said restrictions suggest that, while our result benefits from the combination of provable rewards with AI feedback for reviewing code, further research is needed to address these above challenges. 
This includes expanding language coverage, conducting experiments with more model families, including long CoT models, improving the integration of verifiable tools, developing reliable open-source evaluation methods, and verifying our findings in more diverse environments and deployment contexts, including repository-level code review.

\section*{Ethics Statement}

This research adheres to ethical guidelines and copyright regulations through its use of publicly available datasets augmented with quality annotations from multiple review systems. We obtained all data following appropriate terms of use and ensured proper citation of original sources. To mitigate potential risks of harmful or biased content in LLM-generated pseudo-references, we implemented a multi-stage quality assessment protocol and excluded any potentially problematic examples from our training pipeline. Our human evaluation was conducted with voluntary participants who provided informed consent, and we compensated them fairly according to standard rates. The code review techniques developed in this work aim to improve software quality and security, with possible positive implications for reducing vulnerabilities in deployed systems. Throughout our methodology, we maintained awareness of potential biases in model-generated feedback and took steps to mitigate these effects by using diverse evaluation metrics and balanced test sets across programming languages.

\bibliography{custom}

\begin{thebibliography}{37}
\expandafter\ifx\csname natexlab\endcsname\relax\def\natexlab#1{#1}\fi

\bibitem[{Bai et~al.(2022)Bai, Kadavath, Kundu, Askell, Kernion, Jones, Chen, Goldie, Mirhoseini, McKinnon, Chen, Olsson, Olah, Hernandez, Drain, Ganguli, Li, Tran-Johnson, Perez, Kerr, Mueller, Ladish, Landau, Ndousse, Lukosuite, Lovitt, Sellitto, Elhage, Schiefer, Mercado, DasSarma, Lasenby, Larson, Ringer, Johnston, Kravec, Showk, Fort, Lanham, Telleen-Lawton, Conerly, Henighan, Hume, Bowman, Hatfield-Dodds, Mann, Amodei, Joseph, McCandlish, Brown, and Kaplan}]{bai2022constitutionalaiharmlessnessai}
Yuntao Bai, Saurav Kadavath, Sandipan Kundu, Amanda Askell, Jackson Kernion, Andy Jones, Anna Chen, Anna Goldie, Azalia Mirhoseini, Cameron McKinnon, Carol Chen, Catherine Olsson, Christopher Olah, Danny Hernandez, Dawn Drain, Deep Ganguli, Dustin Li, Eli Tran-Johnson, Ethan Perez, Jamie Kerr, Jared Mueller, Jeffrey Ladish, Joshua Landau, Kamal Ndousse, Kamile Lukosuite, Liane Lovitt, Michael Sellitto, Nelson Elhage, Nicholas Schiefer, Noemi Mercado, Nova DasSarma, Robert Lasenby, Robin Larson, Sam Ringer, Scott Johnston, Shauna Kravec, Sheer~El Showk, Stanislav Fort, Tamera Lanham, Timothy Telleen-Lawton, Tom Conerly, Tom Henighan, Tristan Hume, Samuel~R. Bowman, Zac Hatfield-Dodds, Ben Mann, Dario Amodei, Nicholas Joseph, Sam McCandlish, Tom Brown, and Jared Kaplan. 2022.
\newblock \href {http://arxiv.org/abs/2212.08073} {Constitutional ai: Harmlessness from ai feedback}.

\bibitem[{Chen et~al.(2022)Chen, Zhang, Nguyen, Zan, Lin, Lou, and Chen}]{chen2022codetcodegenerationgenerated}
Bei Chen, Fengji Zhang, Anh Nguyen, Daoguang Zan, Zeqi Lin, Jian-Guang Lou, and Weizhu Chen. 2022.
\newblock \href {http://arxiv.org/abs/2207.10397} {Codet: Code generation with generated tests}.

\bibitem[{Dai et~al.(2025)Dai, Wu, Zheng, Wei, Shi, Jin, Liu, Dun, Huang, and Yan}]{dai2025processsupervisionguidedpolicyoptimization}
Ning Dai, Zheng Wu, Renjie Zheng, Ziyun Wei, Wenlei Shi, Xing Jin, Guanlin Liu, Chen Dun, Liang Huang, and Lin Yan. 2025.
\newblock \href {http://arxiv.org/abs/2410.17621} {Process supervision-guided policy optimization for code generation}.

\bibitem[{Dou et~al.(2024)Dou, Liu, Jia, Xiong, Zhou, Shen, Shan, Huang, Wang, Fan, Xi, Zhou, Ji, Zheng, Zhang, Huang, and Gui}]{dou2024stepcoderimprovecodegeneration}
Shihan Dou, Yan Liu, Haoxiang Jia, Limao Xiong, Enyu Zhou, Wei Shen, Junjie Shan, Caishuang Huang, Xiao Wang, Xiaoran Fan, Zhiheng Xi, Yuhao Zhou, Tao Ji, Rui Zheng, Qi~Zhang, Xuanjing Huang, and Tao Gui. 2024.
\newblock \href {http://arxiv.org/abs/2402.01391} {Stepcoder: Improve code generation with reinforcement learning from compiler feedback}.

\bibitem[{Fard and Mesbah(2013)}]{jsnose}
Amin~Milani Fard and Ali Mesbah. 2013.
\newblock \href {https://doi.org/10.1109/SCAM.2013.6648192} {Jsnose: Detecting javascript code smells}.
\newblock In \emph{2013 IEEE 13th International Working Conference on Source Code Analysis and Manipulation (SCAM)}, pages 116--125.

\bibitem[{Feng et~al.(2025)Feng, Huang, Qu, Zhang, Qin, Zhong, Jiang, Chi, and Zhong}]{feng2025retoolreinforcementlearningstrategic}
Jiazhan Feng, Shijue Huang, Xingwei Qu, Ge~Zhang, Yujia Qin, Baoquan Zhong, Chengquan Jiang, Jinxin Chi, and Wanjun Zhong. 2025.
\newblock \href {http://arxiv.org/abs/2504.11536} {Retool: Reinforcement learning for strategic tool use in llms}.

\bibitem[{Goldie et~al.(2025)Goldie, Mirhoseini, Zhou, Cai, and Manning}]{goldie2025synthetic}
Anna Goldie, Azalia Mirhoseini, Hao Zhou, Irene Cai, and Christopher~D Manning. 2025.
\newblock Synthetic data generation \& multi-step rl for reasoning \& tool use.
\newblock \emph{arXiv preprint arXiv:2504.04736}.

\bibitem[{He et~al.(2025{\natexlab{a}})He, Shi, Zhuo, Treude, Sun, Xing, Du, and Lo}]{he2025codecourtroomllmsnew}
Junda He, Jieke Shi, Terry~Yue Zhuo, Christoph Treude, Jiamou Sun, Zhenchang Xing, Xiaoning Du, and David Lo. 2025{\natexlab{a}}.
\newblock \href {http://arxiv.org/abs/2503.02246} {From code to courtroom: Llms as the new software judges}.

\bibitem[{He et~al.(2025{\natexlab{b}})He, Shi, Zhuo, Treude, Sun, Xing, Du, and Lo}]{he2025code}
Junda He, Jieke Shi, Terry~Yue Zhuo, Christoph Treude, Jiamou Sun, Zhenchang Xing, Xiaoning Du, and David Lo. 2025{\natexlab{b}}.
\newblock From code to courtroom: Llms as the new software judges.
\newblock \emph{arXiv preprint arXiv:2503.02246}.

\bibitem[{Hui et~al.(2024)Hui, Yang, Cui, Yang, Liu, Zhang, Liu, Zhang, Yu, Dang et~al.}]{hui2024qwen2_5}
Binyuan Hui, Jian Yang, Zeyu Cui, Jiaxi Yang, Dayiheng Liu, Lei Zhang, Tianyu Liu, Jiajun Zhang, Bowen Yu, Kai Dang, et~al. 2024.
\newblock Qwen2.5-coder technical report.
\newblock \emph{arXiv preprint arXiv:2409.12186}.

\bibitem[{Jaoua et~al.(2025{\natexlab{a}})Jaoua, Sghaier, and Sahraoui}]{jaoua2025combininglargelanguagemodels}
Imen Jaoua, Oussama~Ben Sghaier, and Houari Sahraoui. 2025{\natexlab{a}}.
\newblock \href {http://arxiv.org/abs/2502.06633} {Combining large language models with static analyzers for code review generation}.

\bibitem[{Jaoua et~al.(2025{\natexlab{b}})Jaoua, Sghaier, and Sahraoui}]{jaoua2025combining}
Imen Jaoua, Oussama~Ben Sghaier, and Houari Sahraoui. 2025{\natexlab{b}}.
\newblock Combining large language models with static analyzers for code review generation.
\newblock \emph{arXiv preprint arXiv:2502.06633}.

\bibitem[{Jaoua et~al.(2025{\natexlab{c}})Jaoua, Sghaier, and Sahraoui}]{combining_LLM_with_static_analyzer}
Imen Jaoua, Oussama~Ben Sghaier, and Houari Sahraoui. 2025{\natexlab{c}}.
\newblock Combining large language models with static analyzers for code review generation.
\newblock \emph{arXiv preprint arXiv:2502.06633}.

\bibitem[{Kingma and Ba(2017)}]{adam_paper}
Diederik~P. Kingma and Jimmy Ba. 2017.
\newblock \href {http://arxiv.org/abs/1412.6980} {Adam: A method for stochastic optimization}.

\bibitem[{Le et~al.(2022)Le, Wang, Gotmare, Savarese, and Hoi}]{codeRL}
Hung Le, Yue Wang, Akhilesh~Deepak Gotmare, Silvio Savarese, and Steven C.~H. Hoi. 2022.
\newblock \href {http://arxiv.org/abs/2207.01780} {Coderl: Mastering code generation through pretrained models and deep reinforcement learning}.

\bibitem[{Lee et~al.(2024)Lee, Phatale, Mansoor, Mesnard, Ferret, Lu, Bishop, Hall, Carbune, Rastogi, and Prakash}]{rlaif}
Harrison Lee, Samrat Phatale, Hassan Mansoor, Thomas Mesnard, Johan Ferret, Kellie Lu, Colton Bishop, Ethan Hall, Victor Carbune, Abhinav Rastogi, and Sushant Prakash. 2024.
\newblock \href {http://arxiv.org/abs/2309.00267} {Rlaif vs. rlhf: Scaling reinforcement learning from human feedback with ai feedback}.

\bibitem[{Li et~al.(2025)Li, Xue, Zhang, Yang, Zhang, Wang, Yu, Hui, Lin, and Liu}]{li2025startselftaughtreasonertools}
Chengpeng Li, Mingfeng Xue, Zhenru Zhang, Jiaxi Yang, Beichen Zhang, Xiang Wang, Bowen Yu, Binyuan Hui, Junyang Lin, and Dayiheng Liu. 2025.
\newblock \href {http://arxiv.org/abs/2503.04625} {Start: Self-taught reasoner with tools}.

\bibitem[{Li et~al.(2022)Li, Lu, Guo, Duan, Jannu, Jenks, Majumder, Green, Svyatkovskiy, Fu, and Sundaresan}]{codereviewerpaper}
Zhiyu Li, Shuai Lu, Daya Guo, Nan Duan, Shailesh Jannu, Grant Jenks, Deep Majumder, Jared Green, Alexey Svyatkovskiy, Shengyu Fu, and Neel Sundaresan. 2022.
\newblock \href {http://arxiv.org/abs/2203.09095} {Automating code review activities by large-scale pre-training}.

\bibitem[{Lin et~al.(2025)Lin, Liu, Gao, Thongtanunam, and Treude}]{lin2025codereviewqa}
Hong~Yi Lin, Chunhua Liu, Haoyu Gao, Patanamon Thongtanunam, and Christoph Treude. 2025.
\newblock Codereviewqa: The code review comprehension assessment for large language models.
\newblock \emph{arXiv preprint arXiv:2503.16167}.

\bibitem[{Loshchilov and Hutter(2019)}]{adamw}
Ilya Loshchilov and Frank Hutter. 2019.
\newblock Decoupled weight decay regularization.
\newblock In \emph{7th International Conference on Learning Representations, {ICLR} 2019, New Orleans, LA, USA, May 6-9, 2019}. OpenReview.net.

\bibitem[{McAleese et~al.(2024)McAleese, Pokorny, Uribe, Nitishinskaya, Trebacz, and Leike}]{mcaleese2024llmcritics}
Nat McAleese, Rai~Michael Pokorny, Juan Felipe~Ceron Uribe, Evgenia Nitishinskaya, Maja Trebacz, and Jan Leike. 2024.
\newblock Llm critics help catch llm bugs.
\newblock \emph{arXiv preprint arXiv:2407.00215}.

\bibitem[{Mroueh(2025)}]{RLVR}
Youssef Mroueh. 2025.
\newblock \href {http://arxiv.org/abs/2503.06639} {Reinforcement learning with verifiable rewards: Grpo's effective loss, dynamics, and success amplification}.

\bibitem[{Naik et~al.(2024)Naik, Alenius, Fried, and Rose}]{crscore_paper}
Atharva Naik, Marcus Alenius, Daniel Fried, and Carolyn Rose. 2024.
\newblock \href {http://arxiv.org/abs/2409.19801} {Crscore: Grounding automated evaluation of code review comments in code claims and smells}.

\bibitem[{{OpenAI}(2024)}]{gpt4o-mini}
{OpenAI}. 2024.
\newblock \href {https://openai.com/index/gpt-4o-mini-advancing-cost-efficient-intelligence/} {Gpt-4o mini: advancing cost-efficient intelligence}.

\bibitem[{OpenAI et~al.(2024)OpenAI, Achiam, Adler, Agarwal, Ahmad, Akkaya, Aleman, Almeida, Altenschmidt, Altman, Anadkat, Avila, Babuschkin, Balaji, Balcom, Baltescu, Bao, Bavarian, Belgum, Bello, Berdine, Bernadett-Shapiro, Berner, Bogdonoff, Boiko, Boyd, Brakman, Brockman, Brooks, Brundage, Button, Cai, Campbell, Cann, Carey, Carlson, Carmichael, Chan, Chang, Chantzis, Chen, Chen, Chen, Chen, Chen, Chess, Cho, Chu, Chung, Cummings, Currier, Dai, Decareaux, Degry, Deutsch, Deville, Dhar, Dohan, Dowling, Dunning, Ecoffet, Eleti, Eloundou, Farhi, Fedus, Felix, Fishman, Forte, Fulford, Gao, Georges, Gibson, Goel, Gogineni, Goh, Gontijo-Lopes, Gordon, Grafstein, Gray, Greene, Gross, Gu, Guo, Hallacy, Han, Harris, He, Heaton, Heidecke, Hesse, Hickey, Hickey, Hoeschele, Houghton, Hsu, Hu, Hu, Huizinga, Jain, Jain, Jang, Jiang, Jiang, Jin, Jin, Jomoto, Jonn, Jun, Kaftan, Łukasz Kaiser, Kamali, Kanitscheider, Keskar, Khan, Kilpatrick, Kim, Kim, Kim, Kirchner, Kiros, Knight, Kokotajlo, Łukasz Kondraciuk,
  Kondrich, Konstantinidis, Kosic, Krueger, Kuo, Lampe, Lan, Lee, Leike, Leung, Levy, Li, Lim, Lin, Lin, Litwin, Lopez, Lowe, Lue, Makanju, Malfacini, Manning, Markov, Markovski, Martin, Mayer, Mayne, McGrew, McKinney, McLeavey, McMillan, McNeil, Medina, Mehta, Menick, Metz, Mishchenko, Mishkin, Monaco, Morikawa, Mossing, Mu, Murati, Murk, Mély, Nair, Nakano, Nayak, Neelakantan, Ngo, Noh, Ouyang, O'Keefe, Pachocki, Paino, Palermo, Pantuliano, Parascandolo, Parish, Parparita, Passos, Pavlov, Peng, Perelman, de~Avila Belbute~Peres, Petrov, de~Oliveira~Pinto, Michael, Pokorny, Pokrass, Pong, Powell, Power, Power, Proehl, Puri, Radford, Rae, Ramesh, Raymond, Real, Rimbach, Ross, Rotsted, Roussez, Ryder, Saltarelli, Sanders, Santurkar, Sastry, Schmidt, Schnurr, Schulman, Selsam, Sheppard, Sherbakov, Shieh, Shoker, Shyam, Sidor, Sigler, Simens, Sitkin, Slama, Sohl, Sokolowsky, Song, Staudacher, Such, Summers, Sutskever, Tang, Tezak, Thompson, Tillet, Tootoonchian, Tseng, Tuggle, Turley, Tworek, Uribe, Vallone,
  Vijayvergiya, Voss, Wainwright, Wang, Wang, Wang, Ward, Wei, Weinmann, Welihinda, Welinder, Weng, Weng, Wiethoff, Willner, Winter, Wolrich, Wong, Workman, Wu, Wu, Wu, Xiao, Xu, Yoo, Yu, Yuan, Zaremba, Zellers, Zhang, Zhang, Zhao, Zheng, Zhuang, Zhuk, and Zoph}]{openai2024gpt4}
OpenAI, Josh Achiam, Steven Adler, Sandhini Agarwal, Lama Ahmad, Ilge Akkaya, Florencia~Leoni Aleman, Diogo Almeida, Janko Altenschmidt, Sam Altman, Shyamal Anadkat, Red Avila, Igor Babuschkin, Suchir Balaji, Valerie Balcom, Paul Baltescu, Haiming Bao, Mohammad Bavarian, Jeff Belgum, Irwan Bello, Jake Berdine, Gabriel Bernadett-Shapiro, Christopher Berner, Lenny Bogdonoff, Oleg Boiko, Madelaine Boyd, Anna-Luisa Brakman, Greg Brockman, Tim Brooks, Miles Brundage, Kevin Button, Trevor Cai, Rosie Campbell, Andrew Cann, Brittany Carey, Chelsea Carlson, Rory Carmichael, Brooke Chan, Che Chang, Fotis Chantzis, Derek Chen, Sully Chen, Ruby Chen, Jason Chen, Mark Chen, Ben Chess, Chester Cho, Casey Chu, Hyung~Won Chung, Dave Cummings, Jeremiah Currier, Yunxing Dai, Cory Decareaux, Thomas Degry, Noah Deutsch, Damien Deville, Arka Dhar, David Dohan, Steve Dowling, Sheila Dunning, Adrien Ecoffet, Atty Eleti, Tyna Eloundou, David Farhi, Liam Fedus, Niko Felix, Simón~Posada Fishman, Juston Forte, Isabella Fulford, Leo
  Gao, Elie Georges, Christian Gibson, Vik Goel, Tarun Gogineni, Gabriel Goh, Rapha Gontijo-Lopes, Jonathan Gordon, Morgan Grafstein, Scott Gray, Ryan Greene, Joshua Gross, Shixiang~Shane Gu, Yufei Guo, Chris Hallacy, Jesse Han, Jeff Harris, Yuchen He, Mike Heaton, Johannes Heidecke, Chris Hesse, Alan Hickey, Wade Hickey, Peter Hoeschele, Brandon Houghton, Kenny Hsu, Shengli Hu, Xin Hu, Joost Huizinga, Shantanu Jain, Shawn Jain, Joanne Jang, Angela Jiang, Roger Jiang, Haozhun Jin, Denny Jin, Shino Jomoto, Billie Jonn, Heewoo Jun, Tomer Kaftan, Łukasz Kaiser, Ali Kamali, Ingmar Kanitscheider, Nitish~Shirish Keskar, Tabarak Khan, Logan Kilpatrick, Jong~Wook Kim, Christina Kim, Yongjik Kim, Jan~Hendrik Kirchner, Jamie Kiros, Matt Knight, Daniel Kokotajlo, Łukasz Kondraciuk, Andrew Kondrich, Aris Konstantinidis, Kyle Kosic, Gretchen Krueger, Vishal Kuo, Michael Lampe, Ikai Lan, Teddy Lee, Jan Leike, Jade Leung, Daniel Levy, Chak~Ming Li, Rachel Lim, Molly Lin, Stephanie Lin, Mateusz Litwin, Theresa Lopez, Ryan
  Lowe, Patricia Lue, Anna Makanju, Kim Malfacini, Sam Manning, Todor Markov, Yaniv Markovski, Bianca Martin, Katie Mayer, Andrew Mayne, Bob McGrew, Scott~Mayer McKinney, Christine McLeavey, Paul McMillan, Jake McNeil, David Medina, Aalok Mehta, Jacob Menick, Luke Metz, Andrey Mishchenko, Pamela Mishkin, Vinnie Monaco, Evan Morikawa, Daniel Mossing, Tong Mu, Mira Murati, Oleg Murk, David Mély, Ashvin Nair, Reiichiro Nakano, Rajeev Nayak, Arvind Neelakantan, Richard Ngo, Hyeonwoo Noh, Long Ouyang, Cullen O'Keefe, Jakub Pachocki, Alex Paino, Joe Palermo, Ashley Pantuliano, Giambattista Parascandolo, Joel Parish, Emy Parparita, Alex Passos, Mikhail Pavlov, Andrew Peng, Adam Perelman, Filipe de~Avila Belbute~Peres, Michael Petrov, Henrique~Ponde de~Oliveira~Pinto, Michael, Pokorny, Michelle Pokrass, Vitchyr~H. Pong, Tolly Powell, Alethea Power, Boris Power, Elizabeth Proehl, Raul Puri, Alec Radford, Jack Rae, Aditya Ramesh, Cameron Raymond, Francis Real, Kendra Rimbach, Carl Ross, Bob Rotsted, Henri Roussez,
  Nick Ryder, Mario Saltarelli, Ted Sanders, Shibani Santurkar, Girish Sastry, Heather Schmidt, David Schnurr, John Schulman, Daniel Selsam, Kyla Sheppard, Toki Sherbakov, Jessica Shieh, Sarah Shoker, Pranav Shyam, Szymon Sidor, Eric Sigler, Maddie Simens, Jordan Sitkin, Katarina Slama, Ian Sohl, Benjamin Sokolowsky, Yang Song, Natalie Staudacher, Felipe~Petroski Such, Natalie Summers, Ilya Sutskever, Jie Tang, Nikolas Tezak, Madeleine~B. Thompson, Phil Tillet, Amin Tootoonchian, Elizabeth Tseng, Preston Tuggle, Nick Turley, Jerry Tworek, Juan Felipe~Cerón Uribe, Andrea Vallone, Arun Vijayvergiya, Chelsea Voss, Carroll Wainwright, Justin~Jay Wang, Alvin Wang, Ben Wang, Jonathan Ward, Jason Wei, CJ~Weinmann, Akila Welihinda, Peter Welinder, Jiayi Weng, Lilian Weng, Matt Wiethoff, Dave Willner, Clemens Winter, Samuel Wolrich, Hannah Wong, Lauren Workman, Sherwin Wu, Jeff Wu, Michael Wu, Kai Xiao, Tao Xu, Sarah Yoo, Kevin Yu, Qiming Yuan, Wojciech Zaremba, Rowan Zellers, Chong Zhang, Marvin Zhang, Shengjia
  Zhao, Tianhao Zheng, Juntang Zhuang, William Zhuk, and Barret Zoph. 2024.
\newblock \href {http://arxiv.org/abs/2303.08774} {Gpt-4 technical report}.

\bibitem[{Ouyang et~al.(2022)Ouyang, Wu, Jiang, Almeida, Wainwright, Mishkin, Zhang, Agarwal, Slama, Ray, Schulman, Hilton, Kelton, Miller, Simens, Askell, Welinder, Christiano, Leike, and Lowe}]{rlhf}
Long Ouyang, Jeff Wu, Xu~Jiang, Diogo Almeida, Carroll~L. Wainwright, Pamela Mishkin, Chong Zhang, Sandhini Agarwal, Katarina Slama, Alex Ray, John Schulman, Jacob Hilton, Fraser Kelton, Luke Miller, Maddie Simens, Amanda Askell, Peter Welinder, Paul Christiano, Jan Leike, and Ryan Lowe. 2022.
\newblock \href {http://arxiv.org/abs/2203.02155} {Training language models to follow instructions with human feedback}.

\bibitem[{Pecorelli et~al.(2022)Pecorelli, Lujan, Lenarduzzi, Palomba, and De~Lucia}]{Pecorelli2022-bp}
Fabiano Pecorelli, Savanna Lujan, Valentina Lenarduzzi, Fabio Palomba, and Andrea De~Lucia. 2022.
\newblock On the adequacy of static analysis warnings with respect to code smell prediction.
\newblock \emph{Empir. Softw. Eng.}, 27(3):64.

\bibitem[{Rafailov et~al.(2024)Rafailov, Sharma, Mitchell, Ermon, Manning, and Finn}]{dpo}
Rafael Rafailov, Archit Sharma, Eric Mitchell, Stefano Ermon, Christopher~D. Manning, and Chelsea Finn. 2024.
\newblock \href {http://arxiv.org/abs/2305.18290} {Direct preference optimization: Your language model is secretly a reward model}.

\bibitem[{Shao et~al.(2024)Shao, Wang, Zhu, Xu, Song, Bi, Zhang, Zhang, Li, Wu, and Guo}]{shao2024deepseekmathpushinglimitsmathematical}
Zhihong Shao, Peiyi Wang, Qihao Zhu, Runxin Xu, Junxiao Song, Xiao Bi, Haowei Zhang, Mingchuan Zhang, Y.~K. Li, Y.~Wu, and Daya Guo. 2024.
\newblock \href {http://arxiv.org/abs/2402.03300} {Deepseekmath: Pushing the limits of mathematical reasoning in open language models}.

\bibitem[{Sharma et~al.(2024)Sharma, Keh, Mitchell, Finn, Arora, and Kollar}]{sharma2024criticalevaluationaialign}
Archit Sharma, Sedrick~Scott Keh, Eric Mitchell, Chelsea Finn, Kushal Arora, and Thomas Kollar. 2024.
\newblock A critical evaluation of ai feedback for aligning large language models.
\newblock \emph{Advances in Neural Information Processing Systems}, 37:29166--29190.

\bibitem[{Sharma(2024)}]{designite}
Tushar Sharma. 2024.
\newblock \href {https://doi.org/10.1145/3643991.3644881} {Multi-faceted code smell detection at scale using designitejava 2.0}.
\newblock In \emph{Proceedings of the 21st International Conference on Mining Software Repositories}, MSR '24, page 284–288, New York, NY, USA. Association for Computing Machinery.

\bibitem[{Sharma et~al.(2022)Sharma, Kechagia, Georgiou, Tiwari, Vats, Moazen, and Sarro}]{sharma2022surveymachinelearningtechniques}
Tushar Sharma, Maria Kechagia, Stefanos Georgiou, Rohit Tiwari, Indira Vats, Hadi Moazen, and Federica Sarro. 2022.
\newblock \href {http://arxiv.org/abs/2110.09610} {A survey on machine learning techniques for source code analysis}.

\bibitem[{Shojaee et~al.(2023)Shojaee, Jain, Tipirneni, and Reddy}]{shojaee2023executionbasedcodegenerationusing}
Parshin Shojaee, Aneesh Jain, Sindhu Tipirneni, and Chandan~K. Reddy. 2023.
\newblock \href {http://arxiv.org/abs/2301.13816} {Execution-based code generation using deep reinforcement learning}.

\bibitem[{Tufano et~al.(2022)Tufano, Masiero, Mastropaolo, Pascarella, Poshyvanyk, and Bavota}]{tufano}
Rosalia Tufano, Simone Masiero, Antonio Mastropaolo, Luca Pascarella, Denys Poshyvanyk, and Gabriele Bavota. 2022.
\newblock Using pre-trained models to boost code review automation.
\newblock In \emph{Proceedings of the 44th international conference on software engineering}, pages 2291--2302.

\bibitem[{Wang et~al.(2025{\natexlab{a}})Wang, Guo, Gao, Fan, Chong, and Xia}]{wang2025llmsreplacehumanevaluators}
Ruiqi Wang, Jiyu Guo, Cuiyun Gao, Guodong Fan, Chun~Yong Chong, and Xin Xia. 2025{\natexlab{a}}.
\newblock \href {http://arxiv.org/abs/2502.06193} {Can llms replace human evaluators? an empirical study of llm-as-a-judge in software engineering}.

\bibitem[{Wang et~al.(2025{\natexlab{b}})Wang, Guo, Gao, Fan, Chong, and Xia}]{wang2025llmjudge}
Ruiqi Wang, Jiyu Guo, Cuiyun Gao, Guodong Fan, Chun~Yong Chong, and Xin Xia. 2025{\natexlab{b}}.
\newblock Can llms replace human evaluators? an empirical study of llm-as-a-judge in software engineering.
\newblock \emph{arXiv preprint arXiv:2502.06193}.

\bibitem[{Zhang et~al.(2023)Zhang, Chen, Liu, Liao, Gong, Yu, Li, and Wang}]{zhang2023unifying}
Ziyin Zhang, Chaoyu Chen, Bingchang Liu, Cong Liao, Zi~Gong, Hang Yu, Jianguo Li, and Rui Wang. 2023.
\newblock Unifying the perspectives of nlp and software engineering: A survey on language models for code.
\newblock \emph{arXiv preprint arXiv:2311.07989}.

\end{thebibliography}
\appendix
\label{sec:appendix}
\section{Methodology Details}
\label{sec:appendix_method}

This appendix contains additional details on the implementation of our CRScore++ method.

\subsection{Code Smell Detector Details}
\label{app:smell_detectors}

\subsubsection{Python (In-domain)}
We use PyScent for Python code smell detection, focusing on structural and maintainability issues. Table~\ref{tab:pyscent} details the key smells detected.

\begin{table*}[!tbh]
\centering
\resizebox{\textwidth}{!}{%
\begin{tabular}{@{}ll@{}}
\toprule
\textbf{Smell Name} & \textbf{Description} \\ 
\midrule
Long Method & Methods exceeding 50 lines, hindering readability and increasing difficulty in testing and maintenance \\
Long Parameter List & Methods with >6 parameters, indicating excessive coupling and potential confusion during method invocation \\
Long Branch & Conditional structures with excessive nesting or length, creating cognitive burden and bug susceptibility \\
Many Attributes & Classes with >15 attributes, suggesting violation of Single Responsibility Principle and excessive state complexity \\
Many Methods & Classes with >20 methods, indicating potential violation of cohesion principles and bloated class responsibilities \\
Shotgun Surgery & When a single change requires modifications across multiple classes, creating maintenance hazards and update risks \\
Low Class Cohesion & Classes where member variables and methods have minimal conceptual relationships, indicating weak internal design \\
Complex Code & Blocks with cyclomatic complexity >15, creating paths difficult to fully test and comprehend \\
Long Lambda & Anonymous functions exceeding reasonable size, subverting their intended use for simple, concise operations \\
Long List Comprehension & List comprehensions exceeding 80 characters, compromising the readability benefits of this Python idiom \\
\bottomrule
\end{tabular}
}
\caption{Python code smells detected by PyScent}
\label{tab:pyscent}
\end{table*}

\subsubsection{Java (Out-of-domain)}
DesigniteJava provides comprehensive smell detection for Java. Tables~\ref{tab:java_design} and~\ref{tab:java_impl} list key detected smells.

\begin{table*}[!htb]
\centering
\resizebox{\textwidth}{!}{\begin{tabular}{ll}
\toprule
\textbf{Design Smell} & \textbf{Description} \\
\midrule
Broken Modularization & Poor separation of components that belong together, violating packaging principles and logical boundaries \\
Cyclic Dependency & Circular references between packages creating tight coupling that complicates testing and independent deployment \\
Hub-like Modularization & Excessive centralization where one package depends on many others, creating potential bottlenecks \\
Deep Hierarchy & Inheritance trees deeper than 5 levels, increasing complexity and reducing comprehensibility of class behavior \\
Multipath Hierarchy & Multiple inheritance paths to the same superclass, leading to potential method resolution ambiguity \\
\bottomrule
\end{tabular}}
\caption{Java design smells (DesigniteJava)}
\label{tab:java_design}
\end{table*}

\begin{table*}[!htb]
\centering
\resizebox{\textwidth}{!}{\begin{tabular}{ll}
\toprule
\textbf{Implementation Smell} & \textbf{Description} \\
\midrule
Complex Method & Methods with cyclomatic complexity >10, creating numerous execution paths difficult to fully test \\
Long Parameter List & Methods requiring more than 5 parameters, indicating potential design problems and complicating API usage \\
Magic Number & Unnamed numerical constants embedded in code, hampering maintainability and self-documentation \\
Empty Catch & Exception blocks that catch errors but take no recovery action, potentially hiding important failure conditions \\
Long Identifier & Variable or method names exceeding 25 characters, reducing readability despite intentions for clarity \\
\bottomrule
\end{tabular}}
\caption{Java implementation smells (DesigniteJava)}
\label{tab:java_impl}
\end{table*}

\subsubsection{JavaScript (Out-of-domain)}
Our custom JSNOSE-based analyzer detects these key smells enlisted in Table \ref{tab:js_smells}

\begin{table*}[!htb]
\centering
\resizebox{\textwidth}{!}{\begin{tabular}{ll}
\toprule
\textbf{Smell} & \textbf{Description} \\
\midrule
Long Function & Functions exceeding 30 lines, creating maintenance challenges in a language often used for UI interactions \\
Excessive Globals & More than 5 global variables in a file, risking namespace pollution and implicit coupling between modules \\
Nested Callbacks & More than 3 levels of callback nesting, creating "callback hell" that impedes code comprehension \\
JS/HTML/CSS Coupling & Direct manipulation of DOM or styles from JS code, violating separation of concerns in frontend development \\
Unused Code & Functions or variables declared but never referenced, adding unnecessary complexity and maintenance overhead \\
Closure Smells & Improperly managed closures that can lead to memory leaks or unexpected variable capture behaviors \\
Refused Bequest & Inheritance relationships where subclasses ignore significant portions of parent functionality \\
\bottomrule
\end{tabular}}
\caption{JavaScript code smells (Custom JSNOSE)}
\label{tab:js_smells}
\end{table*}

\subsection{Teacher Model Dataset Creation Prompt}
\label{app:sft_dataset_creation_prompt}

The system prompt is as follows:
\begin{CustomVerbatim}
You are an exceptionally intelligent coding assistant that consistently delivers accurate and reliable responses to user instructions. You are also experienced in software development and write concise, comprehensive, and relevant comments while reviewing code changes.
\end{CustomVerbatim}

The task-specific, final review comment generation with step by step analysis prompt is as follows:
\begin{CustomVerbatim}
TASK PROMPT: Your task is to look at the given git diff that represents a Python code change as well as linter feedback and code smells detected in the file. You need to write a brief, concise, comprehensive, and relevant review that touches upon all the important topics in the code change as well as all the code smells and linter feedback relevant to that code change.

Code Change:
{code_change}

Code Smells:
{code_smell_detector_messages}

Linter Messages:
{linter_messages}

Now, you should think step by step about all the code changes, code smells, and linter outputs and then decide which things to include. Only pick the smells or linter feedback corresponding to the code change. Also, try to cover aspects like bugs, code security, code readability, maintainability, memory consumption, performance, good and bad design patterns, and efficiency introduced in the code change, which might be missed by these tools. Put your step by step analysis in a section titled "### Step-by-Step Analysis:".

Generate your final review in a section titled "### Final Review:" so that it is easy to parse from your response. The output should only have "### Step-by-Step Analysis:"(should be pointwise) and "### Final Review:" (this should be a single paragraph) sections.

\end{CustomVerbatim}

\subsection{SFT Training Prompt}
\label{app:training_prompt}

The system prompt is same as before. The task specific prompt is given below:
\begin{CustomVerbatim}
Your task is to look at the given git diff that represents a Python code change. You need to write a brief, concise, comprehensive, and relevant review that touches upon all the important topics, code smells, vulnerabilities, and issues in the code change.

Code Change:
{code_change}

Now, you should think step-by-step about all the code changes and identify code smells, issues that would be flagged by a linter, security vulnerabilities, etc. Also, try to cover aspects like bugs, code security, code readability, maintainability, memory consumption, performance, good and bad design patterns, and efficiency introduced in the code change. Put your step-by-step analysis in a section titled "### Step-by-Step Analysis:".

Generate your final review in a section titled "### Final Review:" so that it is easy to parse from your response. The output should only have "### Step-by-Step Analysis:" (should be pointwise) and "### Final Review:" (this should be a single paragraph) sections
\end{CustomVerbatim}


\subsection{Hyperparameters for Training Models}
We train the \texttt{Qwen/Qwen2.5-Coder-3B-Instruct} and \texttt{Qwen/Qwen2.5-Coder-7B-Instruct}, with flash attention, random seed of 42, evaluation every 100 steps (eval\_steps), max length padding, train-val split of 0.1 (10\% data used for validation), batch size of 2, maximum training sequence length of 1024 and 2 training epochs.
The training is done on a single A100, 80 GB GPU for 3B and double A100, 80 GB using hugginface recipes\footnote{\url{https://github.com/huggingface/alignment-handbook/tree/main/recipes}}

\section{Experimental Details}
\label{sec:appendix_experimental_details}

Further experimental details like guidelines for human study annotation of code review quality.

\subsection{LLM-as-a-judge Prompt}
\label{app:llm_as_a_judge_prompt}
\begin{CustomVerbatim}
Your task is to look at a given git diff that represents a Python code change, linter feedback and code smells detected in the code 
change, and a corresponding review comment about the diff. You need to rate how concise, comprehensive, and relevant a review is and whether it touches upon all the important topics, code smells, vulnerabilities, and issues in the code change.

Code Change:
{code_change}

Code Smells:
{code_smell_detector_messages}

Linter Messages:
{linter_messages}

Review Comment:
{review_comment}

You should first generate a step-by-step list of all the topics the review should cover like code smells, issues that would be flagged by a linter, security vulnerabilities, etc. Also, the review should cover aspects like bugs, code security, code readability, maintainability, memory consumption, performance, good and bad design patterns, and efficiency introduced in the code change. Put your analysis under a section titled “### Topics to be Covered:”. 

After generating the list above you should again think step-by-step about the given review comment and whether it addresses these topics and put it under a section called "### Step-by-Step Analysis of Review Comment:". Then based on your step-by-step analysis you should generate a score ranging from 1 (minimum value) to 5 (maximum value) each about how comprehensive, concise, and relevant a review is. A review getting a score of 5 on comprehensiveness addresses nearly all the points in the “### Topics to be Covered:” section while a review scoring 1 addresses none of them. A review getting a score of 5 on conciseness only covers the topics in the  “### Topics to be Covered:” section without wasting time on off-topic information while a review getting a score of 1 is entirely off-topic. Finally, a review scoring 5 on relevance is both concise and comprehensive while a review scoring 1 is neither concise nor comprehensive, effectively making relevance a combined score of conciseness and comprehensiveness. You should give your final rating in a section titled “### Final Scores:”. If a review scores 4 on comprehensiveness, 2 on conciseness, and 3 on relevance give the final scores as shown below (please follow the exact format).

### Final Scores:
```
{"comprehensiveness": 4, "conciseness": 2, "relevance": 3}
```

Now start your analysis starting with the “### Topics to be Covered:”, followed by "### Step-by-Step Analysis of Review Comment:" and ending with the “### Final Scores:”.

### Topics to be Covered:
{topics_to_be_covered}
\end{CustomVerbatim}

\subsection{CoT Evaluation Prompt}
\label{app:cot_evaluation_prompt}

\begin{CustomVerbatim}
Your task is to look at a given git diff that represents a code change, linter feedback, and code smells detected in the code change. You will also be given some “step-by-step analysis” done by a Large Language Model (LLM) that was asked to review the code change which supposedly does linter like and code smell analysis as well. You need to use the LLM’s analysis of the code change along two dimensions: 1) accuracy and 2) coverage. For the “accuracy” dimension rate on a scale of 1 to 5, how factually accurate the step-by-step analysis of the LLM is, given the linter messages and code smell detection, which can be taken as “ground truth”. Meanwhile, the “coverage” rates on a scale of 1 to 5 how many of the linter messages and code smells are covered by the step-by-step analysis.

Code Change:
{code_change}

Code Smells:
{code_smell_detector_messages}

Linter Messages:
{linter_messages}

Step by Step Analysis:
{chain_of_thought}

You should give your final rating in a section titled “### Final Scores:”. If the step-by-step analysis obtains an accuracy of 4 and coverage of 2 with respect to the code smells and linter messages then you should output the results as shown below (please follow the exact format).

### Final Scores:
```
{"accuracy": 4, "coverage": 2}
```

Now provide your rating of the “Step by Step Analysis”:

### Final Scores:


\end{CustomVerbatim}

\subsection{Dataset Statistics}
\label{sec:dataset_stats}
Our experimental pipeline utilized multiple datasets across different training and evaluation phases. For supervised fine-tuning (SFT), we selected 20,888 Python code examples from the CodeReviewer dataset which is publicly available with the Apache 2.0 license\footnote{\url{https://huggingface.co/microsoft/codereviewer}}. We systematically annotated these examples with static analysis signals by generating Ruff linter feedback and PyScent code smell detections for each code patch. Using GPT-4o-mini, we produced structured reviews containing both step-by-step analysis and concise final review comments, accompanied by a comprehensive "topics to be covered" list that served as pseudo-references for subsequent evaluation.
For the preference optimization (DPO) phase, we sampled a subset of 5,000 instances from the SFT dataset. For each instance, we generated 20 diverse candidate reviews using our SFT-trained models with varying temperature settings. These candidates were scored by GPT-4o-mini based on comprehensiveness, conciseness, and relevance metrics, using the previously generated "topics to be covered" as evaluation criteria. We selected pairs with the largest score differentials ($\Delta\geq 2$) for creating the accept/reject pairs required for DPO training.
To validate our approach through human evaluation, we curated a balanced test set comprising 33 Python, 34 Java, and 34 JavaScript instances. Four experienced software developers evaluated the review comments generated by the four model variants described in Section~\ref{sec:human_study}. To ensure reliability, we asked annotators to evaluate an overlapping subset (10\% of total annotations) to measure inter-annotator agreement. The resulting Cohen's kappa score of 0.7 indicates substantial agreement among evaluators, confirming the consistency of our human assessment methodology.

\subsection{Computational Budget}

We utilized significant computational resources throughout our experimental pipeline to train models and generate evaluation data. For model training, we employed NVIDIA A100 80GB GPUs for all variants of the Qwen models described in Section~\ref{Chapter: 6_experimental}. The SFT stage required approximately 8 hours of training (2 epochs) on the 20,888 Python instances from the CodeReviewer dataset, while the DPO stage required an additional 6 hours on a subset of 5,000 instances.
For dataset preparation and evaluation, we made approximately 40,000 API calls to GPT-4o-mini: 20,000 calls for generating "topics to be covered" lists and another 20,000 for producing chain-of-thought analysis and review comments. The LLM-as-a-judge evaluation methodology required additional API calls proportional to our test set sizes: approximately 1,000 calls for Python (evaluating 5 models), and 1,600 calls for Java and JavaScript combined (evaluating 10 models).
Our training configuration employed the AdamW optimizer with a learning rate of 2e-5, a batch size of 8, and a weight decay of 0.01. For the DPO training phase, we used a β value of 0.1 to balance preference adherence against regularization from the reference model. All experiments were conducted with half-precision (fp16) to optimize memory usage while maintaining numerical stability. The total computational budget for our experiments, including training, inference, and evaluation, amounted to approximately 450 GPU hours on A100 hardware.

\subsection{Codebook for Rating Code Review Quality}
\label{sec:codebook_for_rating_code_review_quality}
For the human evaluation, raters were presented with model-generated code reviews and the corresponding code diffs. Unlike CRScore which focused on pseudo-references, our CRScore++ evaluation focused directly on the quality of the review comments in relation to a predefined list of "topics to be covered" for each code change. These topics were systematically generated by GPT-4o-mini and included issues flagged by static analysis tools (linters and code smell detectors) as well as potential security, performance, and maintainability concerns related to the code change.
Each rater was provided with the original code diff, the generated "topics to be covered" list, and access to the complete source files before and after the changes. This comprehensive context enabled raters to make informed judgments about review quality across three dimensions: comprehensiveness, conciseness, and relevance. Reviews were rated on a 5-point Likert scale, with specific scoring guidelines provided to ensure evaluation consistency.

\begin{table*}[!tbhp]
\centering
\resizebox{2\columnwidth}{!}{\begin{tabular}{@{}lrl@{}}
\toprule
Dimension & Score & Rule of thumb \\ \midrule
\multirow{5}{*}{Conciseness} & 1 & None of the review is related to the code change or identified topics. \\
 & 2 & Some of the review is related to the identified topics and static analysis feedback. \\
 & 3 & Roughly half of the review addresses relevant topics without excessive digression. \\
 & 4 & Most of the review effectively addresses identified topics without unnecessary content. \\
 & 5 & The entire review precisely addresses identified topics with optimal brevity. \\ \midrule
\multirow{5}{*}{Comprehensiveness} & 1 & Review fails to address any identified topics or static analysis findings. \\
 & 2 & Review covers at least one code smell, linter warning, or important topic. \\
 & 3 & Review addresses approximately half of the identified topics. \\
 & 4 & Review covers most identified topics, including code smells and linter warnings. \\
 & 5 & Review comprehensively addresses virtually all identified topics and static analysis findings. \\ \midrule
Relevance & & \begin{tabular}[c]{@{}l@{}}Relevance score represents the balance between comprehensiveness and\\ conciseness, and must fall between those two scores.\end{tabular} \\
 & & \begin{tabular}[c]{@{}l@{}}When one dimension significantly underperforms, the relevance score\\ should be weighted toward that limiting factor.\\ \\ For example: A review that identifies all relevant code smells (high\\ comprehensiveness) but includes substantial off-topic discussion (low\\ conciseness) should have a relevance score closer to its conciseness rating.\end{tabular} \\ \bottomrule
\end{tabular}}
\caption{Scoring guidelines for evaluating code review quality in CRScore++. These rules provide a framework for consistent assessment across multiple annotators, focusing on how effectively reviews address topics identified by static analysis tools and other code quality considerations.}
\label{tab:crscore_plus_quality_scores}
\end{table*}

\subsection{Rating Review Quality using Topics}
\label{sec:rating_review_quality_using_topics}
 Our method employes a "topics to be covered" approach to standardize review evaluation. Before evaluation, we generated a comprehensive list of topics for each code change, encompassing:
\begin{itemize}
\item Code smells detected by language-specific tools (PyScent, DesigniteJava, JSNOSE)
\item Linter warnings from static analyzers (Ruff, PMD, custom JavaScript linter)
\item Security vulnerabilities and potential edge cases
\item Readability, maintainability, and performance considerations
\end{itemize}
Raters evaluated code reviews from four different model configurations on a 5-point Likert scale for each dimension. Table~\ref{tab:crscore_plus_quality_scores} shows our scoring guidelines. For comprehensiveness, scores ranged from 1 (covering none of the topics) to 5 (addressing virtually all topics). For conciseness, scores ranged from 1 (mostly irrelevant content) to 5 (focused entirely on relevant topics). Relevance combined both aspects, representing overall review quality.

\subsection{Annotator Information}

We hired five graduate students with experience in writing Java, Python, and Javascript code. The annotation process took approximately three weeks, with each annotator evaluating 33-34 examples per programming language across four different model configurations.
Before evaluation began, annotators participated in a comprehensive two-hour training session on code smell recognition, static analysis fundamentals, and quality assessment guidelines. We conducted calibration exercises where annotators rated the same set of examples and discussed discrepancies to align their scoring approaches. We iteratively refined our guidelines until achieving the reported inter-annotator agreement of 0.7 (Cohen's kappa).
All annotators signed informed consent forms acknowledging that their evaluations would be used for research purposes and potentially published in anonymized form. The annotators were not informed about which model produced which review to prevent potential biases toward specific systems. For the compensation,
the annotators who took up this task were paid in 30 USD in Amazon gift cards after successful completion of the study.

\subsection{Ethics Review for Data Collection}
The human evaluation protocol for CRScore++ was reviewed and approved by the institutional review board (IRB) of the university the authors are affiliated with.

\section{More Results And Analysis}
\label{sec:appendix_results_and_analysis}
This section presents additional statistical analyses that complement the model performance results discussed in the main paper. We conduct a series of Wilcoxon Signed Rank tests to establish the statistical significance of differences observed between various model configurations. All tests use an appropriate significance level (α = 0.01 or α = 0.005) to ensure robustness of our findings.

\subsection{Model Size Comparisons}
Table~\ref{tab:size-comparison} examines the impact of model size by comparing 3B and 7B variants of each configuration. Interestingly, not all dimensions show significant differences between model sizes. For the "CR dataset Python only" and Tool Guided approaches, conciseness scores show no statistically significant differences (p>0.01), suggesting that increasing model size does not necessarily improve review brevity. However, comprehensiveness and relevance show significant improvements with the larger model size in most configurations, particularly for our Stage 1 and Stage 2 models. This indicates that additional parameters enable the model to capture more complex relationships between code smells, linter warnings, and their implications.

\subsection{Comparison Across 3B Model Variants}
The detailed comparisons in Table~\ref{tab:qwen3b-variants-significance} reveal several important patterns among 3B model variants. First, models trained on the CodeReviewer dataset alone consistently underperform across all metrics compared to other approaches (p<0.005), reinforcing our hypothesis that traditional supervised fine-tuning on human-written reviews is insufficient for high-quality review generation. The progression from Zero Shot to Tool Guided to Stage 1 to Stage 2 shows statistically significant improvements in both comprehensiveness and relevance, confirming the incremental benefits of our training pipeline.
Notably, the transition from Zero Shot to Tool Guided shows significant improvements in comprehensiveness and relevance (p<0.005) without significantly affecting conciseness, suggesting that tool integration provides substantial benefits without introducing verbosity. The comparison between Stage 1 and Stage 2 models reveals significant differences across all dimensions (p<0.005), with Stage 2 showing enhanced comprehensiveness and relevance at a small but significant cost to conciseness.

\subsection{Comparison Across 7B Model Variants}
Table~\ref{tab:qwen7b-variants-significance} presents similar comparisons for 7B model variants. The patterns largely mirror those seen in 3B models, but with some notable differences. Unlike the 3B comparison, there is no significant difference between the Zero Shot and Tool Guided approaches in the 7B setting across any metric (p>0.005). This suggests that larger models may already capture much of the knowledge provided by static analysis tools, making explicit tool integration less impactful.
Both Stage 1 and Stage 2 models demonstrate highly significant improvements over baseline approaches (p<0.005) across most metrics. The comparison between Stage 1 and Stage 2 shows significant differences in all dimensions, confirming that preference optimization consistently enhances review quality independent of model size, albeit with a modest reduction in conciseness. This pattern is consistent with our finding that DPO training guides models toward more thorough coverage of code quality issues, sometimes at the expense of brevity.
These statistical analyses provide strong evidence for the effectiveness of our CRScore++ training methodology across different model sizes and highlight the specific contributions of each component in our approach. The consistent patterns of improvement across both 3B and 7B models demonstrate the scalability of our approach and its potential application to even larger model architectures.

\begin{table*}[htbp]
\centering
\resizebox{2\columnwidth}{!}{%
\begin{tabular}{llcl}
\hline
\multicolumn{1}{c}{\textbf{Language}} &
  \multicolumn{1}{c}{\textbf{Code Change}} &
  \textbf{Model Type} &
  \multicolumn{1}{c}{\textbf{Model's Code Change Review}} \\ \hline
Java &
  \begin{tabular}[c]{@{}l@{}}@@ -53,7 +53,7 @@ public class ProtocGapic \\ PluginGeneratorTest \{ \\ ~~~~~model.getFiles.stream \\ ~~~~~.mapProtoFile::getProto \\ ~~~~~.collectCollectors.toList) \\ ~~~~~// Only the file to generate \\ ~~~~~.addFileToGenerate"multiple\_services.proto" \\ -~~~~.setParameter"language=java" \\ +~~~~.setParameter"language=java,transport=grpc" \\ ~~~~~.build;\end{tabular} &
  Qwen 3B Stage 1 &
  \begin{tabular}[c]{@{}l@{}}I noticed that a new parameter "transport=grpc" \\ has been added to the setParameter() method.\\ This change introduces a potential code smell \\ as it increases the dependency on external \\ function calls. The test may now have a \\ "shotgun smell" due to high coupling with \\ external components. I recommend adding\\ documentation for this parameter.\end{tabular} \\ \hline
JavaScript &
  \begin{tabular}[c]{@{}l@{}}@@ -124,6 +124,12 @@ function validateConfigconfig \\ ~~~~~if !config.hasOwnProperty'timeout' \{ \\ ~~~~~~~~config.timeout = 30000; \\ ~~~~~\} \\ +~~~~if !config.hasOwnProperty'retries' \{ \\ +~~~~~~~~config.retries = 3; \\ +~~~~\} \\ +~~~~if config.retries < 0 \{ \\ +~~~~~~~~throw new Error'Retries cannot be negative'; \\ +~~~~\} \\ ~~~~~return config; \\ ~~~\}\end{tabular} &
  Qwen 3B Stage 1 &
  \begin{tabular}[c]{@{}l@{}}The PR adds retry functionality to the config\\ validation. I see a potential code smell here\\ with the repetitive property checking pattern.\\ Linter would flag this as a violation of the\\ DRY principle. Consider refactoring to use a\\ helper function for property validation. Also,\\ the Error message could be more descriptive\\ for debugging purposes.\end{tabular} \\ \hline
Java &
  \begin{tabular}[c]{@{}l@{}}@@ -85,7 +85,7 @@ class RequestHandler \{ \\ ~~~private void processRequestRequest req \{ \\ ~~~~~try \{ \\ -~~~~~~logger.log"Processing request " + req.getId; \\ +~~~~~~logger.debug"Processing request " + req.getId; \\ ~~~~~\} catch Exception e \{ \\ ~~~~~~~handleExceptione; \\ ~~~~~\}\end{tabular} &
  Qwen 3B Stage 2 DPO &
  \begin{tabular}[c]{@{}l@{}}This change modifies the logging level from\\ generic log to debug level. This addresses a\\ common code smell related to inappropriate\\ logging levels. The linter would have flagged\\ this as a violation of logging best practices.\\ However, there's still a performance concern\\ with string concatenation in logging statements\\ which should be addressed using parameterized\\ logging instead.\end{tabular} \\ \hline
JavaScript &
  \begin{tabular}[c]{@{}l@{}}@@ -45,7 +45,9 @@ class DataProcessor \{ \\ ~~~processdata \{ \\ ~~~~~if !data return null; \\ -~~~~return this.transformdata; \\ +~~~~const validated = this.validatedata; \\ +~~~~if !validated throw new Error'Invalid data'; \\ +~~~~return this.transformvalidated; \\ ~~~\} \\ ~~~// ... rest of the class\end{tabular} &
  Qwen 3B Stage 2 DPO &
  \begin{tabular}[c]{@{}l@{}}The PR adds data validation before processing.\\ This addresses a code smell where input\\ validation was missing. A linter would have\\ flagged the original code for potential null\\ reference errors. However, I notice another\\ code smell: throwing errors in the middle of\\ processing logic creates fragmented control flow.\\ Consider returning an error result instead\\ following the Either pattern for better error\\ handling.\end{tabular} \\ \hline
\end{tabular}%
}
\caption{Comparison of model responses across different stages for out-of-domain languages (Java and JavaScript). The examples highlight how our models effectively identify code smells and linter issues in unfamiliar languages.}
\label{tab:model-comparison}
\end{table*}

\begin{table*}[htbp]
\centering

\resizebox{2\columnwidth}{!}{\begin{tabular}{l 
                cc   cc   cc}
\toprule
Comparison 
  & \multicolumn{2}{c}{Comprehensiveness} 
  & \multicolumn{2}{c}{Conciseness} 
  & \multicolumn{2}{c}{Relevance} \\

  & p-value & statistic 
  & p-value & statistic 
  & p-value & statistic \\

\midrule
Qwen 3B (CR dataset\textsubscript{\texttt{Python only}}) vs. Qwen 7B (CR dataset\textsubscript{\texttt{Python only}}) 
  & \textbf{p<0.01} & 7.50e+06 
  & p>0.01 & 8.07e+06 
  & \textbf{p<0.01} & 7.50e+06 \\

Qwen 3B Zero Shot vs.\ Qwen 7B Zero Shot
  & p>0.01 & 8.34e+06 
  & \textbf{p<0.01 }& 8.67e+06 
  & \textbf{p<0.01} & 8.49e+06 \\

Qwen 3B (Tool Guided)  
   vs.\ Qwen 7B (Tool Guided)
  & p>0.01 & 8.17e+06 
  & \textbf{p<0.01} & 8.57e+06 
  & p>0.01 & 8.25e+06 \\

Qwen 3B Stage 1 vs.\ Qwen 7B Stage 1
  & \textbf{p<0.01} & 8.71e+06 
  & \textbf{p<0.01} & 7.73e+06 
  & \textbf{p<0.01} & 8.64e+06 \\

Qwen 3B Stage 2 vs.\ Qwen 7B Stage 2
  & \textbf{p<0.01} & 7.32e+06 
  & \textbf{p<0.01} & 8.75e+06 
  & \textbf{p<0.01} & 7.28e+06 \\

\bottomrule
\end{tabular}}
\caption{Statistical significance Wilcoxon Signed Rank tests performed among Qwen 3B and 7B variants on comprehensiveness, conciseness, and relevance (significance level $\alpha = 0.01$). We highlight statistically significant observations.}
\label{tab:size-comparison}
\end{table*}

\begin{table*}[htbp]
\centering
\resizebox{2\columnwidth}{!}{%
\begin{tabular}{l  cc  cc  cc}
\toprule
Comparison 
  & \multicolumn{2}{c}{Comprehensiveness} 
  & \multicolumn{2}{c}{Conciseness} 
  & \multicolumn{2}{c}{Relevance} \\

  & p-value & statistic 
  & p-value & statistic 
  & p-value & statistic \\

\midrule
Qwen 3B (CR dataset\textsubscript{\texttt{Python only}}) vs.\ Qwen 3B Zero Shot  
  & \textbf{p<0.005} & 1.48e+06 
  & \textbf{p<0.005} & 5.65e+06 
  & \textbf{p<0.005} & 2.57e+06 \\

Qwen 3B (CR dataset\textsubscript{\texttt{Python only}}) vs.\ Qwen 3B (Tool Guided)  
  & \textbf{p<0.005} & 1.70e+06 
  & \textbf{p<0.005} & 5.62e+06 
  & \textbf{p<0.005} & 2.82e+06 \\

Qwen 3B (CR dataset\textsubscript{\texttt{Python only}}) vs.\ Qwen 3B Stage 1  
  & \textbf{p<0.005} & 4.07e+05 
  & \textbf{p<0.005} & 5.69e+06 
  & \textbf{p<0.005} & 1.20e+06 \\

Qwen 3B (CR dataset\textsubscript{\texttt{Python only}}) vs.\ Qwen 3B Stage 2  
  & \textbf{p<0.005} & 2.56e+05 
  & \textbf{p<0.005} & 7.38e+06 
  & \textbf{p<0.005} & 8.48e+05 \\

Qwen 3B Zero Shot vs.\ Qwen 3B (Tool Guided) 
  & \textbf{p<0.005} & 8.40e+06 
  & p>0.005 & 8.07e+06 
  & \textbf{p<0.005} & 8.42e+06 \\

Qwen 3B Zero Shot vs.\ Qwen 3B Stage 1  
  & \textbf{p<0.005} & 4.77e+06 
  & p>0.005 & 8.17e+06 
  & \textbf{p<0.005} & 5.36e+06 \\

Qwen 3B Zero Shot vs.\ Qwen 3B Stage 2  
  & \textbf{p<0.005} & 3.39e+06 
  & \textbf{p<0.005} & 1.00e+07 
  & \textbf{p<0.005} & 4.02e+06 \\

Qwen 3B (Tool Guided) vs.\ Qwen 3B Stage 1  
  & \textbf{p<0.005} & 4.60e+06 
  & p>0.005 & 8.21e+06 
  & \textbf{p<0.005} & 5.11e+06 \\

Qwen 3B (Tool Guided) vs.\ Qwen 3B Stage 2  
  & \textbf{p<0.005} & 3.28e+06 
  & \textbf{p<0.005} & 1.01e+07 
  & \textbf{p<0.005} & 3.82e+06 \\

Qwen 3B Stage 1 vs.\ Qwen 3B Stage 2  
  & \textbf{p<0.005} & 6.30e+06 
  & \textbf{p<0.005} & 9.98e+06 
  & \textbf{p<0.005} & 6.51e+06 \\

\bottomrule
\end{tabular}%
}
\caption{Statistical significance Wilcoxon Signed Rank tests performed among Qwen 3B variants on comprehensiveness, conciseness, and relevance (significance level $\alpha = 0.005$). We highlight statistically significant observations.}
\label{tab:qwen3b-variants-significance}
\end{table*}

\begin{table*}[ht]
\centering
\resizebox{2\columnwidth}{!}{%
\begin{tabular}{l  cc  cc  cc}
\toprule
Comparison 
  & \multicolumn{2}{c}{Comprehensiveness} 
  & \multicolumn{2}{c}{Conciseness} 
  & \multicolumn{2}{c}{Relevance} \\

  & p-value & statistic 
  & p-value & statistic 
  & p-value & statistic \\

\midrule
Qwen 7B (CR dataset\textsubscript{\texttt{Python only}}) vs.\ Qwen 7B Zero Shot  
  & \textbf{p<0.005} & 2.82e+06 
  & \textbf{p<0.005} & 6.45e+06 
  & \textbf{p<0.005} & 4.03e+06 \\

Qwen 7B (CR dataset\textsubscript{\texttt{Python only}}) vs.\ Qwen 7B (Tool Guided)
  & \textbf{p<0.005} & 2.87e+06 
  & \textbf{p<0.005} & 6.31e+06 
  & \textbf{p<0.005} & 4.05e+06 \\

Qwen 7B (CR dataset\textsubscript{\texttt{Python only}}) vs.\ Qwen 7B Stage 1  
  & \textbf{p<0.005} & 7.67e+05 
  & \textbf{p<0.005} & 5.31e+06 
  & \textbf{p<0.005} & 1.74e+06 \\

Qwen 7B (CR dataset\textsubscript{\texttt{Python only}}) vs.\ Qwen 7B Stage 2  
  & \textbf{p<0.005} & 7.20e+05 
  & p>0.005 & 8.01e+06 
  & \textbf{p<0.005} & 1.31e+06 \\

Qwen 7B Zero Shot vs.\ Qwen 7B (Tool Guided)  
  & p>0.005 & 8.15e+06 
  & p>0.005 & 7.99e+06 
  & p>0.005 & 8.13e+06 \\

Qwen 7B Zero Shot vs.\ Qwen 7B Stage 1  
  & \textbf{p<0.005} & 5.37e+06 
  & \textbf{p<0.005} & 7.27e+06 
  & \textbf{p<0.005} & 5.74e+06 \\

Qwen 7B Zero Shot vs.\ Qwen 7B Stage 2  
  & \textbf{p<0.005} & 3.19e+06 
  & \textbf{p<0.005} & 9.80e+06 
  & \textbf{p<0.005} & 3.63e+06 \\

Qwen 7B (Tool Guided) vs.\ Qwen 7B Stage 1  
  & \textbf{p<0.005} & 5.36e+06 
  & \textbf{p<0.005} & 7.41e+06 
  & \textbf{p<0.005} & 5.73e+06 \\

Qwen 7B (Tool Guided) vs.\ Qwen 7B Stage 2  
  & \textbf{p<0.005} & 3.23e+06 
  & \textbf{p<0.005} & 9.96e+06 
  & \textbf{p<0.005} & 3.64e+06 \\

Qwen 7B Stage 1 vs.\ Qwen 7B Stage 2  
  & \textbf{p<0.005} & 5.10e+06 
  & \textbf{p<0.005} & 1.09e+07 
  & \textbf{p<0.005} & 5.32e+06 \\

\bottomrule
\end{tabular}%
}
\caption{Statistical significance Wilcoxon Signed Rank tests performed among Qwen 7B variants on comprehensiveness, conciseness, and relevance (significance level $\alpha = 0.005$). We highlight statistically significant observations.}
\label{tab:qwen7b-variants-significance}
\end{table*}



\end{document}